\documentclass[camera,letterpaper,nomarginnotes,nonarrowgutter]{jpaper} 

\usepackage{cite}
\usepackage{amsmath,amssymb,amsfonts}
\usepackage{algorithmic}
\usepackage{graphicx}
\usepackage{textcomp}
\usepackage[bookmarks=true,breaklinks=true,letterpaper=true,colorlinks,linkcolor=black,citecolor=black,urlcolor=black]{hyperref}
\usepackage{balance} 
\usepackage{fancyhdr}

\def\BibTeX{{\rm B\kern-.05em{\sc i\kern-.025em b}\kern-.08em
    T\kern-.1667em\lower.7ex\hbox{E}\kern-.125emX}}
    
\usepackage[linesnumbered,ruled]{algorithm2e}
\usepackage{algorithmic}
\usepackage{graphicx}
\usepackage{textcomp}
\usepackage{xcolor}
\usepackage{fancyhdr}
\usepackage{xspace}
\usepackage{datetime} 
\usepackage[normalem]{ulem}
\usepackage{booktabs}
\usepackage{multirow}
\usepackage{setspace}
\usepackage{siunitx}
\usepackage{marginnote}
\usepackage{pifont}
\usepackage{enumitem}
\newcommand{\boldone}{\ding{202}}
\newcommand{\boldtwo}{\ding{203}}
\newcommand{\boldthree}{\ding{204}}
\newcommand{\boldfour}{\ding{205}}
\newcommand{\boldfive}{\ding{206}}
\newcommand{\boldsix}{\ding{207}}
\newcommand{\boldseven}{\ding{208}}
\newcommand{\boldeight}{\ding{209}}

\usepackage[colorinlistoftodos,prependcaption,textsize=small]{todonotes}
\usepackage{xargs}                    
\usepackage{titlesec}
\usepackage{titletoc}
\usepackage{clipboard}
\usepackage[all]{nowidow}
\usepackage[framemethod=tikz]{mdframed}
\usepackage[most]{tcolorbox}
\tcbuselibrary{breakable}
\tcbuselibrary{hooks}
\usepackage[font={small,bf}]{caption}
\usepackage{subcaption}
\usepackage{titlesec}
\usepackage{dblfloatfix}

\newcommand{\onur}[1]{{#1}}

\newcommand\X[0]{PiDRAM\xspace}

\newcommand{\xmark}{\ding{55}}%

\newcommand\outline[1]{\textcolor{orange}{\textbf{Outline:} #1}}

\newif\ifrebuttal
\rebuttalfalse
\newif\iftacorev
\tacorevfalse

\ifrebuttal
\newcommand{\reva}[1]{\textcolor{red}{#1}}
\definecolor{dark-green}{rgb}{0.00, 0.45, 0.00}
\newcommand{\revb}[1]{\textcolor{dark-green}{#1}}
\definecolor{goldbutdark}{rgb}{0.85, 0.65, 0.12}
\newcommand{\revd}[1]{\textcolor{goldbutdark}{#1}}
\newcommand{\revf}[1]{\textcolor{blue}{#1}}
\newcommand{\revcommon}[1]{\textcolor{purple}{#1}}
\definecolor{pink-hot}{rgb}{0.98, 0.40, 0.78}
\newcommand{\reve}[1]{\textcolor{pink-hot}{#1}}
\else
\newcommand{\reva}[1]{#1}
\newcommand{\revb}[1]{#1}
\newcommand{\revd}[1]{#1}
\newcommand{\revf}[1]{#1}
\newcommand{\revcommon}[1]{#1}
\newcommand{\reve}[1]{#1}
\fi

\iftacorev

\definecolor{dark-green}{rgb}{0.00, 0.45, 0.00}

\definecolor{goldbutdark}{rgb}{0.85, 0.65, 0.12}

\definecolor{pink-hot}{rgb}{0.98, 0.40, 0.78}

\fi

\newif\ifsubmission
\submissiontrue

\ifsubmission

\newcommand{\new}[1]{#1}
\newcommand{\newnew}[1]{#1}
\newcommand{\omi}[1]{#1}
\newcommand{\omu}[1]{#1}
\newcommand{\revdel}[1]{}

\definecolor{ups-truck}{rgb}{0.53, 0.28, 0.21}
\newcommand{\atb}[1]{\textcolor{black}{#1}}

\definecolor{dgreen}{rgb}{0.00, 0.75, 0.00}
\newcommand{\jgl}[1]{[\textit{{\color{black}JGL: #1}}]}
\newcommand{\juan}[1]{{\color{black}#1}}

\definecolor{dblue}{rgb}{0.00, 0.00, 1.00}

\else
\definecolor{ups-truck}{rgb}{0.53, 0.28, 0.21}
\newcommand{\atb}[1]{\textcolor{ups-truck}{#1}}

\definecolor{dgreen}{rgb}{0.00, 0.75, 0.00}
\newcommand{\jgl}[1]{[\textit{{\color{dgreen}JGL: #1}}]}
\newcommand{\juan}[1]{{\color{dgreen}#1}}

\definecolor{dblue}{rgb}{0.00, 0.00, 1.00}

\fi

\newcommand{\affilETH}[0]{\textsuperscript{\S}}
\newcommand{\affilETU}[0]{\textsuperscript{$\dagger$}}

\definecolor{lightblue}{rgb}{0.980, 0.956, 0.623}

\let\oldmarginnote\marginnote

\renewcommand{\marginnote}[2][rectangle,draw,fill=blue!40,rounded corners]{%
        \oldmarginnote{%
        \tikz \node at (0,0) [#1]{#2};}%
        }
        
\definecolor{lightyellow}{rgb}{0.980, 0.956, 0.623}

\newcommand{\boxbegin} {
	\begin{tcolorbox}[enhanced, frame hidden, colback=gray!50, breakable]
}

\newcommand{\boxend} {
	\end{tcolorbox}
}

\newcommand{\yboxbegin} {
	\begin{tcolorbox}[breakable, enhanced, frame hidden, colback=yellow!50]
}

\newcommand{\yboxend} {
	\end{tcolorbox}
}

\mdfdefinestyle{graybox}{
    splittopskip=0,%
    splitbottomskip=0,%
    frametitleaboveskip=0,
    frametitlebelowskip=0,
    skipabove=0,%
    skipbelow=0,%
    leftmargin=0,%
    rightmargin=0,%
    innertopmargin=0mm,%
    innerbottommargin=0mm,%
    roundcorner=1mm,%
    backgroundcolor=lightblue,
    hidealllines=true}
    
\mdfdefinestyle{graybox2}{
    splittopskip=0,%
    splitbottomskip=0,%
    frametitleaboveskip=0,
    frametitlebelowskip=0,
    skipabove=0,%
    skipbelow=0,%
    leftmargin=0,%
    rightmargin=0,%
    innertopmargin=2mm,%
    innerbottommargin=2mm,%
    roundcorner=2mm,%
    backgroundcolor=lightblue,
    hidealllines=true}

\newcommand{\bboxbegin}{
    \begin{mdframed}[style=graybox]
}

\newcommand{\bboxend}{
    \end{mdframed}
}

\newcommand{\yyboxbegin}{
    \begin{mdframed}[style=graybox2]
}

\newcommand{\yyboxend}{
    \end{mdframed}
}

\title{PiDRAM: A Holistic End-to-end FPGA-based Framework\\for \underline{P}rocessing-\underline{i}n-\underline{DRAM}}
\author{
{Ataberk Olgun\affilETH}\qquad%
{Juan G\'omez Luna\affilETH}\qquad
{Konstantinos Kanellopoulos\affilETH}\qquad
{Behzad Salami\affilETH}\qquad\\
{Hasan Hassan\affilETH}\qquad%
{O\=guz Ergin\affilETU}\qquad%
{Onur Mutlu\affilETH}\qquad\vspace{-3mm}\\\\
{\vspace{-3mm}\affilETH \emph{ETH Z{\"u}rich}} \qquad \affilETU \emph{TOBB University of Economics and Technology}
}

\begin{document}

\maketitle

\thispagestyle{plain}
\pagestyle{plain}

\begin{abstract}
\juan{
\new{Commodity DRAM based processing-using-memory (PuM) techniques that are\revdel{ reliably} supported by off-the-shelf DRAM chips present an opportunity for alleviating the data movement bottleneck at low cost.}
However, system integration of \new{these} techniques imposes non-trivial challenges that are yet to \atb{be} solve\atb{d}. 
Potential solutions to the integration challenges require appropriate tools to develop \newnew{any} necessary hardware and software components. 
Unfortunately, current proprietary computing systems, specialized DRAM-testing platforms, or system simulators do not provide the flexibility and/or the holistic system view that is necessary to \newnew{properly evaluate and} deal with \new{the} integration challenges \new{of commodity DRAM based PuM techniques}.}

\juan{We design and develop \X, \atb{the first} flexible end-to-end framework that enables system integration studies and evaluation of real, {commodity DRAM based} PuM techniques. 
\X provides software and hardware \atb{components} to rapidly integrate PuM techniques across the whole system software and hardware stack. 
We implement \X on an FPGA-based RISC-V system.}
\atb{To demonstrate the flexibility and ease of use of \X, we implement \juan{and evaluate} two state-of-the-art {commodity DRAM based} PuM techniques: \new{(i) in-DRAM copy and initialization (RowClone) and (ii) in-DRAM true random number generation (D-RaNGe)}}.
\revdel{First, we implement 
\juan{in-memory copy and initialization. We propose solutions to integration challenges and conduct a detailed end-to-end implementation study.}
\atb{Second, we implement} \juan{a true random number generator in DRAM.}
Our results show that
\juan{the in-memory copy and initialization techniques can} improve the performance of bulk copy operations by 12.6$\times{}$ and bulk initialization operations by 14.6$\times{}$ \juan{on a real system}.
Implementing
\juan{the true random number generator} requires \juan{only} 190 lines of Verilog and 74 lines of C code
\juan{using \X's software and hardware \atb{components}.}} \new{We describe how we solve key integration challenges to make \newnew{such} techniques \newnew{work and be effective} on a real-system prototype, including memory allocation, alignment, and coherence.} \new{We observe that end-to-end RowClone speeds up bulk copy and initialization operations by 14.6$\times{}$ and 12.6$\times{}$, respectively over conventional CPU copy\newnew{, even when coherence is supported with inefficient cache flush operations}.} \new{Over PiDRAM's extensible codebase, integrating both RowClone and D-RaNGe end-to-end on a real RISC-V system prototype takes \newnew{only} 388 lines of Verilog \newnew{code} and 643 lines of C++ code.}
\end{abstract}


\section{Introduction}
\label{sec:introduction}

\revdel{DRAM-based main memory is used in nearly all computing systems as a major component. The growing memory footprints and working-sets of modern workloads require main memory to satisfy three properties. \new{Main memory needs to be} (i) \emph{fast}, so that memory accesses induce small latency costs \juan{and applications can enjoy high memory bandwidth}, (ii) \emph{dense}, so that working-sets of workloads fit into the memory device and \juan{do} not need to be \juan{frequently} brought from \juan{secondary} storage, and (iii) \emph{low-power}, to lower energy consumption, prevent devices from overheating, and increase battery life in mobile computing systems. 
Unfortunately, 
the trends in memory technology development show that it is especially difficult to scale existing memory devices in all three dimensions~\cite{mutlu2020modern}. 
\juan{In this context, DRAM vendors have prioritized memory capacity scaling over latency and bandwidth~\cite{chang.sigmetrics16,lee.hpca13}}. 
As a result, Memory performance improvements have been lagging behind processor performance improvements in recent years and m}Main memory \newnew{is a major performance and energy} bottleneck in computing systems~\cite{mutlu2020modern,ghose2019processing}.
%
%
One way of overcoming the main memory bottleneck is to move computation into/near memory, \juan{a paradigm known as \emph{processing-in-memory} (PiM)~\cite{mutlu2020modern}}. 
\juan{PiM} reduces memory latency between the 
\juan{memory units and the compute units}, enables the 
\juan{compute units} to \atb{exploit the large internal bandwidth within} memory devices,  
and reduces the overall power consumption of the system by eliminating the need for transferring data over power-hungry off-chip interfaces~\cite{mutlu2020modern,ghose2019processing}.

Recent works propose a variety of \juan{PiM} techniques to alleviate the data movement problem. 
\Copy{R4/4}{\juan{One {set} of techniques propose to place compute {logic} \emph{near} memory arrays {(e.g., processing capability in the memory controller{, logic layer of 3D-stacked memory,} or {near} the memory array within the memory chip)} }~\cite{fernandez2020natsa,cali2020genasm,kim.bmc18,ahn.pei.isca15,ahn.tesseract.isca15,boroumand.asplos18,boroumand2019conda,boroumand2016pim,singh2019napel,asghari-moghaddam.micro16,JAFAR,farmahini2015nda,gao.pact15,DBLP:conf/hpca/GaoK16,gu.isca16,hashemi.isca16,cont-runahead,hsieh.isca16,kim.isca16,kim.sc17,liu-spaa17,morad.taco15,nai2017graphpim,pattnaik.pact16,pugsley2014ndc,zhang.hpdc14,zhu2013accelerating,DBLP:conf/isca/AkinFH15,gao2017tetris,drumond2017mondrian,dai2018graphh,zhang2018graphp,huang2020heterogeneous,zhuo2019graphq,syncron}. 
\juan{These techniques are called \emph{processing-near-memory} (PnM)} \new{techniques}~\cite{mutlu2020modern}.
\juan{Another {set of techniques} propose to leverage analog properties of memory {(e.g., SRAM, DRAM, and NVM)} operation to perform computation in different ways {(e.g., leveraging non-deterministic behavior in memory array operation to generate random numbers, performing bitwise operations within the memory array by exploiting analog charge sharing properties of DRAM operation) }\Copy{R4/1}{~\cite{aga.hpca17,eckert2018neural,fujiki2019duality,kang.icassp14,chang.hpca16,seshadri.micro17,seshadri2013rowclone,angizi2019graphide,li.dac16,angizi2018pima,angizi2018cmp,angizi2019dna,levy.microelec14,kvatinsky.tcasii14,shafiee2016isaac,kvatinsky.iccd11,kvatinsky.tvlsi14,gaillardon2016plim,bhattacharjee2017revamp,hamdioui2015memristor,xie2015fast,hamdioui2017myth,yu2018memristive,rezaei2020nom,wang2020figaro,mandelman.ibmjrd02,xin2020elp2im,gao2020computedram,li.micro17,deng.dac2018,kim.hpca18,kim.hpca19,hajinazarsimdram,ali2019memory,ronen2022bitlet,zha2019liquid,testa2016inversion,borghetti2010memristive,intel-loihi,geoffrey2017neuromorphic}.} 
\juan{These techniques are known as \emph{processing-using-memory} (PuM)} \new{techniques}~\cite{mutlu2020modern}.}}



A subset of PuM proposals devise mechanisms that enable computation using DRAM arrays~\cite{seshadri.micro17,seshadri2013rowclone,angizi2019graphide,kim.hpca18,kim.hpca19,gao2020computedram,chang.hpca16,xin2020elp2im,li.micro17,deng.dac2018,hajinazarsimdram,rezaei2020nom,wang2020figaro,ali2019memory}. 
\juan{These mechanisms provide significant performance benefits and energy savings by exploiting the high internal bit-level parallelism of DRAM for (1) {bulk data} copy and initialization operations {\newnew{at row} granularity}~\cite{seshadri2013rowclone,chang.hpca16,rezaei2020nom,wang2020figaro,aga.hpca17}, (2) bitwise operations~\cite{seshadri.micro17,xin2020elp2im,li.dac16,angizi2018pima,Seshadri:2015:ANDOR,seshadri.arxiv16,seshadri.bookchapter17.arxiv,seshadri.thesis16,li.micro17,mandelman.ibmjrd02,angizi2018cmp,angizi2019dna}, (3) arithmetic operations~\cite{levy.microelec14,kvatinsky.tcasii14,aga.hpca17,kang.icassp14,li.micro17,shafiee2016isaac,eckert2018neural,fujiki2019duality,kvatinsky.iccd11,kvatinsky.tvlsi14,gaillardon2016plim,bhattacharjee2017revamp,hamdioui2015memristor,xie2015fast,hamdioui2017myth,yu2018memristive,deng.dac2018,angizi2019graphide}, and (4) security primitives (e.g., true random number generation~\cite{kim.hpca19}, physical unclonable functions~\cite{kim.hpca18,orosa2021codic})}. 
Recent works~\cite{gao2020computedram,kim.hpca19,kim.hpca18} show that \juan{some} of these PuM mechanisms \juan{can already be} reliably supported in contemporary, \juan{off-the-shelf} DRAM chips.\footnote{{We are especially interested in PiM techniques that do \emph{not} require {any} modification to the DRAM chips or the DRAM interface.}} 
Given that DRAM is \newnew{the} 
\juan{dominant \newnew{main} memory technology}, these {commodity DRAM based} PuM techniques\footnote{{Commodity DRAM based PuM techniques are PuM techniques that can already be supported in existing off-the-shelf DRAM chips with\newnew{out \emph{any}} modification to \newnew{DRAM chips or DRAM interfaces}.}} provide a promising way to improve the performance and energy efficiency of existing and future systems at \emph{no additional \juan{DRAM} hardware cost}.

\juan{Integration of these PuM mechanisms in a real system imposes non-trivial challenges that require further research to find appropriate solutions.}
\juan{For example, in-DRAM {bulk data} copy and initialization \newnew{techniques}~\cite{seshadri.micro17,chang.hpca16} require modifications \newnew{to} memory management that affect different parts of the system. 
First, these \newnew{techniques} have specific memory allocation and alignment requirements ({e.g., page-granularity} source and destination operand arrays \newnew{should be} allocated and aligned in the same DRAM subarray) that are \emph{not} satisfied by existing memory allocation primitives {(e.g., \texttt{malloc}~\cite{malloc}, \texttt{posix\_memalign}~\cite{posixmemalign})}. 
Second, in-DRAM copy requires efficient handling of memory coherence, such that the contents of the source operand in DRAM are up-to-date}.
\revdel{\juan{Another example of an integration challenge is the need for bit transposition of in-DRAM computation mechanisms~\cite{hajinazarsimdram,ali2019memory,angizi2019graphide} that employ vertical data layout.}}




\juan{None of these system integration challenges of PuM mechanisms can be {efficiently} studied in existing {general-purpose} computing systems {(e.g., personal computers, cloud computers, embedded systems)}, special-purpose testing platforms {(e.g., SoftMC~\cite{hassan2017softmc})}, or system simulators {(e.g., gem5~\cite{gem5-gpu,GEM5}, Ramulator~\cite{ramulator.github,ramulator}, {Ramulator-PIM~\cite{ramulator-pim},} zsim~\cite{zsim}, \newnew{DAMOVSim~\cite{geraldodamov,oliveira2021damov}}, \new{\newnew{and} other simulators~\cite{zhang2022pim,forlin2022sim2pim,yu2021multipim,xu2019pimsim})}}.} 
\atb{First, many {commodity DRAM based} PuM mechanisms in DRAM rely on non-standard DDRx operation, where timing parameters for DDRx commands} \juan{are violated~\cite{gao2020computedram,kim.hpca18,kim.hpca19}} {(or otherwise new DRAM commands are added, which requires new chip designs and interfaces)}. 
{Existing general-purpose} computing systems do \emph{not} permit dynamically changing DDRx timing parameters, which is required to integrate \juan{these PuM mechanisms} into real systems. 
Second, prior works show that the reliability of {commodity DRAM based} PuM mechanisms is highly dependent on environmental \atb{conditions} such as temperature and voltage fluctuations~\cite{kim.hpca18,kim.hpca19} \newnew{and process variation}. These effects are exacerbated by the non-standard behavior of \atb{PuM} mechanisms in real DRAM devices. 
Although \newnew{special-purpose} testing platforms {(e.g., SoftMC~\cite{hassan2017softmc})} can be used to conduct \juan{reliability studies}, these platforms do \emph{not} model \juan{an end-to-end} computing system, \juan{where system integration of PuM mechanisms can be studied.} 
System simulators {\newnew{(e.g., those aforementioned)}} can model \juan{end-to-end computing systems}. However, 
\juan{they \atb{(i) do \emph{not} model DRAM operation \newnew{that violates} manufacturer-recommended timing \newnew{parameters}, (ii) do \emph{not} have {a way of interfacing with real DRAM chips that {embody undisclosed and unique} characteristics {that ha{ve} implications on how PuM techniques are integrated into real systems} (e.g., {proprietary and chip-specific} DRAM internal address mapping~\cite{cojocar2020susceptible,salp,patel2022case}),} and (iii)} \emph{cannot} support characterization studies} on the reliability \juan{of} \atb{PuM} mechanisms 
\juan{since system simulators do not model environmental conditions \newnew{and process variation}.}


\juan{Our goal is to design and implement a flexible \newnew{real-system} platform that can be used to solve system integration challenges and analyze trade-offs of end-to-end implementations of {commodity DRAM based} PuM mechanisms. To this end, we develop \emph{\textbf{P}rocessing-\textbf{i}n-\textbf{DRAM}} (\X) framework, \atb{the first flexible, end-to-end, and open source framework that enables system integration studies and evaluation of real PuM techniques using real {unmodified} DRAM devices.} 

\X facilitates system integration studies of new {commodity DRAM based} PuM mechanisms by providing \atb{four} customizable hardware and software components that can be used as \atb{a} common basis to enable system support for such mechanisms in real systems.}
\juan{\X contains \atb{two} main \emph{hardware} components. 
First, a custom, easy-to-extend \emph{memory controller} allows for \atb{implementing} new DRAM command sequences that \new{perform PuM operations}. For example, \new{the memory controller can be extended with a single state machine in its hardware description to implement a new DDRx command sequence with \new{user-defined} timing parameters to \newnew{implement a new PuM technique (i.e., perform a new PuM operation)}.}\revdel{ we can program this memory controller to issue DDRx command sequences with violated timing parameters that are needed to implement some PuM mechanisms~\cite{kim.hpca18,kim.hpca19,gao2020computedram}}}
\juan{Second, an \emph{ISA-transparent controller (\textbf{P}uM \textbf{O}perations \textbf{C}ontroller, POC)} supervises PuM execution.} \new{POC exposes the PuM operations to the software components of PiDRAM over a memory-mapped interface to the processor, allowing the programmer to perform PuM operations using the PiDRAM framework by executing \newnew{conventional} LOAD/STORE instructions. The memory-mapped interface allows PiDRAM to be easily ported to systems that implement different instruction set architectures.} 
\juan{\new{PiDRAM contains two main \emph{software} components}. 
First, an \emph{extensible library} allows system designers to implement software support for PuM mechanisms. This library contains customizable functions that \atb{communicate with POC to perform PuM operations.}}
\juan{Second, a custom \emph{supervisor software} \atb{contains the necessary OS primitives (e.g., memory management) to enable end-to-end implementations of {commodity DRAM based} PuM techniques.}} 


\atb{We demonstrate a prototype of \X on an FPGA-based RISC-V system~\cite{asanovic2016rocket}. 
To demonstrate the flexibility and ease of use of \X, we implement two \newnew{prominent} PuM \juan{\newnew{techniques}}: (1) \emph{RowClone}~\cite{seshadri2013rowclone}, \juan{an in-DRAM \omi{data} copy and initialization \newnew{technique}}, and (2) \emph{D-RaNGe}~\cite{kim.hpca19}, \juan{an in-DRAM true random number generat{ion technique} based on activation-latency failure\atb{s}}. 
\juan{In order to support RowClone (Section~\ref{sec:rowclone})}, (i) we \juan{customize} \newnew{the} \X memory controller 
\juan{to} issue carefully-engineered \revdel{valid }sequences of DRAM commands that perform \omi{data} copy \newnew{(and initialization)} operations in DRAM, \juan{and} (ii) we extend the custom supervisor software to implement a new memory management mechanism \juan{that} satisfies the memory allocation \juan{and alignment} requirements of RowClone.} 
\juan{For D-RaNGe (Section~\ref{sec:drange}), \newnew{we extend} \new{(i) \newnew{the} \X{} memory controller with a new state machine that periodically performs DRAM accesses \newnew{with reduced activation latencies} to generate random numbers~\cite{kim.hpca19} and a new hardware \emph{random number buffer} that stores the generated random numbers, and (ii) the custom supervisor software with a function that retrieves the random numbers from the hardware buffer to the user program.}}
%
\juan{Our end-to-end evaluation of (i) RowClone \newnew{demonstrates} up to {14.6$\times{}$} speedup for bulk copy and {12.6$\times{}$} initialization operations over CPU copy (i.e., conventional \texttt{memcpy}), \newnew{even when coherence is satisfied using inefficient cache flush operations,} and (ii) D-RaNGe \newnew{demonstrates} that an end-to-end integration of D-RaNGe can provide true random numbers at high throughput (8.30 Mb/s) and low latency (4-bit random number in \SI{220}{\nano\second})\newnew{, even without any \omi{hardware or software} optimizations}.} \new{Implementing both PuM techniques over the Verilog and C++ codebase provided by PiDRAM \newnew{requires} only 388 lines of Verilog \newnew{code} and 643 lines of C++ code.}

Our contributions are as follows:

\begin{itemize}
  \item We develop \X, the first flexible framework that enables end-to-end integration and evaluation of PuM \juan{mechanisms} using real {unmodified} DRAM chips.
  
  \item We develop a prototype of \X on an FPGA-based platform. To demonstrate the ease-of-use {and evaluation benefits} of \X, we implement two state-of-the-art DRAM-based PuM \juan{mechanisms, RowClone and D-RaNGe,} and evaluate them on \X's prototype {using unmodified DDR3 chips}. 
  
  \item We devise a new memory management mechanism that satisfies the memory allocation \juan{and alignment} requirements of RowClone. We demonstrate that our mechanism enables RowClone end-to-end {in the full system, and} \revcommon{provid{es}} significant performance improvements \revcommon{over traditional CPU-{based} copy and initialization operations (\texttt{memcpy}~\cite{memcpy} and \texttt{calloc}~\cite{calloc}) as demonstrated on our PiDRAM prototype}. 
  
  \item \atb{We implement and evaluate a {state-of-the-art} DRAM-based true random number generat{ion technique} (D-RaNGe). Our implementation provides a {solid foundation} for future work on system integration of DRAM-based PuM security primitives (e.g., PUFs~\cite{talukder2019exploiting,kim.hpca18}, TRNGs~\cite{olgun2021quactrngieee,olgun2021quactrng,talukder2019exploiting}){, implemented using real unmodified DRAM chips}.}
\end{itemize}

\section{Background}

{We provide the relevant background on DRAM organization, DRAM operation and commodity DRAM based PuM techniques. We refer the reader to prior works for more comprehensive background about DRAM organization and operation~\cite{salp,lee.hpca13,donghyuk-ddma,chang.sigmetrics17,ghose2018vampire,patel2017reaper,luo2020clr,ghose2019demystifying,kevinchang-thesis,yoongu-thesis,lee.thesis16,olgun2021quactrng}.}

\subsection{DRAM Background}
\label{sec:background-dram}
\begin{figure}[h]
     \centering
     
     \begin{subfigure}[h]{.50\textwidth}
         \centering
         \includegraphics[width=\textwidth]{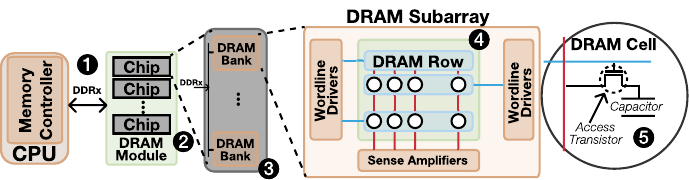}
     \end{subfigure}
     \hfill
     \begin{subfigure}[h]{.45\textwidth}
         \centering
         \includegraphics[width=\textwidth]{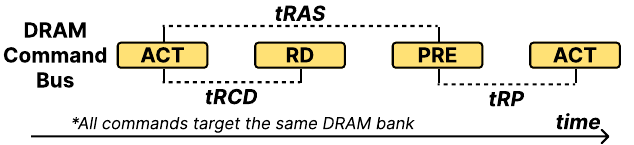}
     \end{subfigure}
    
    \caption{DRAM organization (top). Timing diagram of \new{DRAM} commands (bottom).}
    
    \label{fig:dram-bank-timing-diagram}
\end{figure}

DRAM-based main memory is organized hierarchically. \new{Fig.~\ref{fig:dram-bank-timing-diagram} (top) depicts this organization.} A processor is connected to one or \newnew{more} {memory channels \new{(DDRx in the figure)~\boldone{}}}. Each channel has its own command, address, and data buses. Multiple {memory modules} can be plugged into a single channel. Each module contains several {DRAM chips}\new{~\boldtwo{}. Each chip} contains multiple {\new{DRAM} banks} \new{that can be accessed} independently\new{~\boldthree{}}.\revdel{A set of DDRx standards cluster multiple {banks} in {bank groups}~\cite{jedecDDR4,gddr5}.}
\newnew{D}ata transfers between DRAM memory modules and processors occur at {cache block} granularity. The cache block size is typically 64 bytes in \atb{current} systems.

\new{Inside a DRAM bank, DRAM cells are laid out \newnew{as} a two dimensional array of wordlines \new{(i.e., DRAM rows)} and bitlines \newnew{(i.e., DRAM columns)~\boldfour{}. W}ordlines are depicted in blue and bitlines are depicted in red in Fig.~\ref{fig:dram-bank-timing-diagram}. Wordline drivers drive the wordlines and sense amplifers read the values on the bitlines.} \newnew{A DRAM cell is connected to a bitline via an access transistor\new{~\boldfive{}}}. When enabled, an access transistor allows charge to flow between a DRAM cell and the cell's bitline.

\noindent
\new{\textbf{DRAM Operation.}} When all DRAM rows in a {bank} are closed, DRAM bitlines are precharged to \newnew{a reference voltage level of} {$\frac{V_{DD}}{2}$}. The memory controller sends an activate ($ACT$) command \new{to the DRAM module} to drive a DRAM wordline (i.e., enable a DRAM row). Enabling a DRAM row starts the {charge sharing} process. \newnew{Each DRAM cell connected to the DRAM row starts sharing its charge with its bitline}. This \omi{causes} \newnew{the bitline voltage} to deviate from {$\frac{V_{DD}}{2}$} {(i.e., the charge in the cell perturbs the bitline voltage)}. The \new{sense amplifier} sense\newnew{s} the deviation in the bitline and amplif\newnew{ies} the voltage of the bitline either to {$V_{DD}$} or to $0$. \newnew{As such}, \newnew{an ACT command} copies one DRAM row to the \new{sense amplifiers} (i.e., row buffer). The memory controller can send READ\newnew{/}WRITE commands to transfer data from/to {the sense amplifier array}. {Once the memory controller needs to access another DRAM row, t}he memory controller can close the {enabled} DRAM row by sending a precharge (PRE) command on the command bus. The PRE command first disconnects DRAM cells from their bitlines by disabling the enabled wordline and then precharges the bitlines to {$\frac{V_{DD}}{2}$}.

\noindent
\new{\textbf{DRAM Timing Parameters.}} DRAM datasheets specify a set of timing parameters that define the minimum time window between valid combinations of DRAM commands~\cite{lee.hpca15, kevinchang-thesis, chang.sigmetrics16, kim2018solar}. The memory controller must wait for tRCD, tRAS, and tRP nanoseconds between successive ACT $\rightarrow$ RD, ACT $\rightarrow$ ACT, and PRE $\rightarrow$ ACT commands, respectively {(Figure~\ref{fig:dram-bank-timing-diagram}, bottom)}. Prior works show that these timing parameters can be violated (e.g., successive ACT $\rightarrow$ RD commands may be issued with a shorter time window than tRCD) to improve DRAM access latency~\cite{lee.hpca15, chang.sigmetrics16, kevinchang-thesis, lee.sigmetrics17, kim2018solar}, implement physical unclonable functions~\cite{kim.hpca18,talukder2019exploiting,orosa2021codic}, \newnew{generate true random numbers~\cite{olgun2021quactrng,olgun2021quactrngieee,kim.hpca19},} copy data~\cite{seshadri2013rowclone,gao2020computedram}, and perform bitwise AND/OR operations~\cite{seshadri.micro17,seshadri.thesis16,seshadri.arxiv16,seshadri.bookchapter17.arxiv,gao2020computedram} in commodity DRAM devices.

\noindent
\textbf{\new{{DRAM Internal Address Mapping.}}}
\Copy{R2/1C}{{DRAM manufacturers use DRAM-internal address mapping schemes~\cite{salp,cojocar2020susceptible,patel2022case} to translate from logical (e.g., row, bank, column) DRAM addresses \newnew{that are used by the memory controller} to physical DRAM addresses \newnew{that are internal to the DRAM chip} (e.g., the \newnew{physical} position of a DRAM row within the chip). These schemes allow (i) post-manufacturing row repair techniques to map erroneous DRAM rows to redundant DRAM rows and (ii) DRAM manufacturers to organize DRAM internals in a cost-efficient \newnew{and reliable} way~\cite{khan.dsn16,vandegoor2002address}. DRAM-internal address mapping schemes can be substantially different across different DRAM chips~\cite{barenghi2018software,cojocar2020susceptible,horiguchi1997redundancy,itoh2013vlsi,keeth2001dram,khan.dsn16,khan.micro17,kim-isca2014,lee.sigmetrics17,liu.isca13,orosaYaglikci2021deeper,saroiu2022price,patel2020bit,patel2022case}. Thus, \newnew{consecutive} logical DRAM row addresses might not point to physical DRAM rows in the same subarray.}}


\subsection{PuM Techniques}

\label{sec:background_pudram}
Prior work proposes a variety of in-DRAM computation mechanisms (i.e., PuM techniques) that (i) have great potential to improve system performance and energy efficiency~\cite{chang.hpca16,seshadri.micro17, hajinazarsimdram,seshadri2013rowclone,seshadri2020indram,seshadri.bookchapter17.arxiv,seshadri.bookchapter17,seshadri.arxiv16,Seshadri:2015:ANDOR,angizi2019graphide, ferreira2021pluto} 
\juan{or} (ii) can provide low-cost security primitives~\cite{talukder2019prelatpuf,talukder2019exploiting,kim.hpca19,kim.hpca18,olgun2021quactrngieee,orosa2021codic}. A subset of these in-DRAM computation mechanisms are demonstrated on real DRAM chips~\cite{gao2020computedram,kim.hpca19, kim.hpca18, talukder2019exploiting,olgun2021quactrngieee,orosa2021codic}. {We describe the \newnew{major relevant} prior works \omi{briefly}}: 




\noindent
\newnew{\textbf{RowClone~\cite{seshadri2013rowclone}} \omi{is} a low-cost DRAM architecture that can perform bulk data movement operations (e.g., copy, initialization) inside DRAM chips \omi{at high performance and low energy}.}

\noindent
\newnew{\textbf{Ambit~\cite{Seshadri:2015:ANDOR, seshadri.arxiv16, seshadri.micro17, seshadri.bookchapter17, seshadri2020indram}} \omi{is} a new DRAM substrate that can perform \omi{(i)} bitwise majority \omi{(and thus bitwise AND/OR)} \omi{operations} across three DRAM rows by simultaneously activating three DRAM rows \omi{and (ii) bitwise NOT} operation\omi{s} \omi{on a DRAM row} using 2-transistor 1-capacitor DRAM cells~\cite{kang2009one,lu2015improving}.}

\noindent
\textbf{ComputeDRAM~\cite{gao2020computedram}} demonstrates in-DRAM copy {(previously proposed by RowClone~\cite{seshadri2013rowclone})} and bitwise AND/OR operations {(previously proposed by Ambit~\cite{seshadri.micro17})} on real DDR3 chips. ComputeDRAM performs in-DRAM operations by issuing carefully-engineered, valid sequences of DRAM commands {with violated tRAS and tRP timing parameters (i.e., by not obeying manufacturer-recommended timing parameters defined in DRAM \newnew{chip} specifications~\cite{micron2016ddr4})}. By issuing command sequences {with violated timing parameters}, \omi{ComputeDRAM \omi{activates}} two or three DRAM rows in a DRAM bank \newnew{in quick succession} \omi{(i.e., performs two or three row activation\omi{s})}. \newnew{ComputeDRAM leverages (i) two row activation\omi{s} to transfer data between two DRAM rows and (ii) three row activation\omi{s} to perform the majority function in real unmodified DRAM chips.} 
\revdel{ComputeDRAM leverages multiple row activation (i) to transfer data between two DRAM cells, and (ii) to perform {the} majority function across three rows {in real unmodified DRAM chips}.}

\noindent
\textbf{D-RaNGe~\cite{kim.hpca19}} is a {state-of-the-art} high-throughput DRAM-based true random number generat{ion technique}. D-RaNGe leverages the randomness in DRAM activation (tRCD) failures as its entropy source. D-RaNGe extracts random bits from DRAM cells that fail with $50\%$ probability when accessed with a reduced \newnew{(i.e., violated)} tRCD. 
D-RaNGe demonstrates {high-quality true random number generation} on a vast number of real DRAM chips across multiple generations.

\noindent
\newnew{\textbf{QUAC-TRNG~\cite{olgun2021quactrngieee}} demonstrates that four DRAM rows can be activated in a quick succession using an ACT-PRE-ACT command sequence \omi{(called QUAC)} with violated tRAS and tRP timing parameters in real DDR4 DRAM chips. QUAC-TRNG uses QUAC to generate true random numbers at high throughput and low latency.}



\section{Motivation}

\atb{Implementing DRAM-based PuM techniques and integrating them into a real system requires modifications across the hardware and software stack. End-to-end implementations of PuM techniques require proper tools that (i) are flexible, to enable rapid development of PuM techniques and (ii) support real DRAM devices, to correctly observe the effects of reduced DRAM timing operations that are fundamental \atb{to} enabling {commodity DRAM based} PuM in {real unmodified} DRAM devices. Existing {general-purpose} computers, specialized DRAM testing platforms (e.g., \newnew{those aforementioned, Section~\ref{sec:introduction}}) cannot be used to study end-to-end implementations of {commodity DRAM based} PuM techniques.}

\atb{First, implementing new DDRx command sequences that perform PuM operations requires modifications to the memory controller. {Existing general purpose} computers do not support customizations to the memory controller {to dynamically modify manufacturer-recommended DRAM timing parameters}~\cite{lee.hpca15,kim2018solar,chang.sigmetrics16,kim.hpca19,hassan2017softmc}. This hinders the possibility of studying end-to-end implementations of PuM techniques on such platforms. Second, PuM techniques impose data mapping and allocation requirements (Section~\ref{sec:rowclone}) that are not satisfied by current memory management and allocation mechanisms (e.g., malloc~\cite{malloc}). Current OS memory management schemes must be augmented to satisfy these requirements. Existing specialized DRAM testing platforms (e.g., SoftMC~\cite{hassan2017softmc}) do not have system support \newnew{to enable this}. By design, these platforms are not built for system integration. Hence, it is difficult to evaluate system-level mechanisms that enable PuM techniques on DRAM testing platforms. Third, system simulators (i) do \emph{not} model DRAM operation \newnew{that violates} manufacturer-recommended {timing parameters}, (ii) do \emph{not} have a way of interfacing with {real DRAM chips} that {embody undisclosed and unique characteristics that have implications on how PuM techniques are integrated into real systems (e.g., proprietary and chip-specific DRAM internal address mapping~\cite{cojocar2020susceptible,salp,patel2022case})} that influence PuM operations{, and (iii) \emph{cannot} support studies on the reliability of PuM techniques since system simulators do \emph{not} model environmental conditions \newnew{and process variation}.}}
\atb{We summarize the limitations of the relevant experimental platforms \newnew{later in} Table~\ref{table:tools}.}

\atb{Our \textbf{goal} is to develop a flexible \juan{end-to-end} framework that enables rapid \newnew{system} integration of {commodity DRAM based} PuM techniques and facilitates studies on end-to-end \newnew{full-system} implementations of PuM techniques using real DRAM devices. To this end, we develop \X.}


\section{PiDRAM}

\begin{figure*}[!h]
  \centering
  \includegraphics[width=0.9\textwidth]{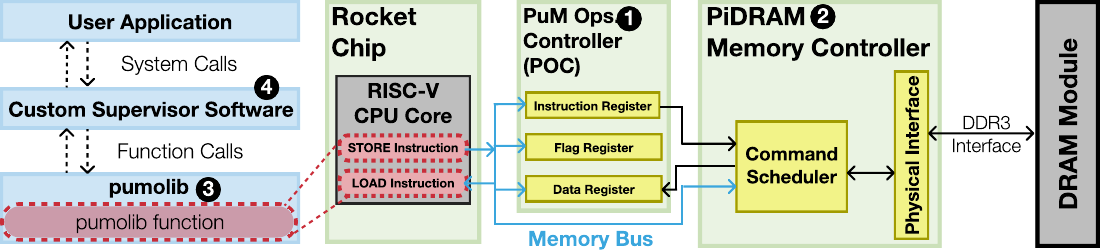}
  \caption{{PiDRAM overview. {Modified h}ardware {(in green)} and software {(in blue)} components. \revd{Unmodified components are in gray.} {A pumolib function executes load and store instructions in the CPU to perform PuM operations (in red).} We use yellow to highlight the key hardware structures that are controlled by the user to perform PuM operations.}}
  \label{fig:pidram-overview}
\end{figure*}

\sloppy
\revdel{\atb{We design the \X framework to} {solve system integration challenges and analyze trade-offs of end-to-end implementations of {commodity DRAM based} PuM techniques}
facilitate end-to-end implementations of {commodity DRAM based} PuM techniques.}
\new{Implementing commodity DRAM based PuM techniques end-to-end requires developing new hardware (HW) and software (SW) components or augmenting existing components with new functionality (e.g., memory allocation for RowClone requires a new memory allocation routine in the OS, Section~\ref{sec:rowclone_alignment}).}
To ease the process of modifying various components across the hardware and software stack to implement new PuM techniques, \X provides key HW and SW components. Figure~\ref{fig:pidram-overview} presents an overview of the HW and SW components of the \X framework. \Copy{R4/8}{{Later in Section~\ref{sec:execution-overview}, we describe the general workflow for executing a PuM operation on PiDRAM.}}
\label{sec:pidram}
\begin{table*}[!b]
  \centering
  \caption{{Pumolib functions}}
  \label{table:pumolib}
  \scriptsize
  \begin{tabular}{@{} lm{15em}m{46em} @{}}
  \toprule
  {\textbf{Function}} &  {\textbf{Arguments}} &  {\textbf{Description}}\\        
  \midrule
  \textbf{set\_timings} & RowClone\_T1, RowClone\_T2, tRCD & Updates CRF registers with the timing parameters used in RowClone (\emph{T1} and \emph{T2}) and D-RaNGe (\emph{tRCD}) operations.\\
  \textbf{rng\_configure} & period, address, bit\_offsets & Updates CRF registers, configuring the random number generator to to access the DRAM cache block at \emph{address} every \emph{period} cycles and collect the bits at \emph{bit\_offsets} from the cache block.\\
  \textbf{copy\_row} & source\_address, destination\_address & Performs a RowClone-Copy operation in DRAM from the \emph{source\_address} to the \emph{destination\_address}.\\
  \textbf{activation\_failure} & address & Induces an activation failure in a DRAM location pointed by the \emph{address}.\\
  \textbf{buf\_\omi{size}} & - & Returns the \newnew{number of random words in} the random number buffer.\\
  \textbf{rand\_dram} & - & Returns 32 bits \newnew{(\omi{i.e.,} random words)} from the random number buffer.\\
  \midrule
  \end{tabular}
\end{table*}

\subsection{Hardware Components}
\label{sec:hardware-components}


{PiDRAM comprises two key hardware components. Both of these components are designed with the goal to provide a flexible and easy to use framework for evaluating PuM techniques.}


\textbf{\ding{182} PuM Operations Controller \newnew{(POC)}.} {POC decodes and executes PiDRAM instructions (e.g., RowClone-Copy~\cite{seshadri2013rowclone} that are used by the programmer to perform PuM operations. POC communicates with the rest of the system over two well{-}defined interfaces. First, it communicates with the CPU over a memory-mapped interface{, where the CPU can send data to or receive data from POC using memory store and load instructions}. The CPU accesses the memory-mapped registers (\emph{instruction}, \emph{data}, and \emph{flag} registers) in POC to execute in-DRAM operations. This improves the portability of the framework and facilitates porting the framework to systems that employ different instruction set architectures. Second, {POC} communicates with the memory controller to perform PuM operations in the DRAM chip over a simple hardware interface. {To do so,} POC (i) requests the memory controller to perform a PuM operation, (ii) waits until the memory controller performs the operation, and (iii) receives the result of the PuM operation from the memory controller. The CPU can read the result of the operation by executing load instructions that target the \emph{data} register in POC.}

\atb{\textbf{\ding{183} Custom Memory Controller.} {PiDRAM's memory controller provides an easy-to-extend basis for {commodity DRAM based} PuM techniques that require issuing DRAM commands with violated timing parameters~\cite{gao2020computedram,kim.hpca19, kim.hpca18, talukder2019exploiting,olgun2021quactrngieee}. 
The memory controller is designed modularly and requires {easy{-}to{-}make} modifications to its scheduler to implement new PuM techniques.} For instance, our modular design enables supporting RowClone operations (Section~\ref{sec:rowclone}) in just 60 lines of Verilog code on top of the baseline custom memory controller's scheduler that implements conventional DRAM operations (e.g., read, write).}

\Copy{R1/6}{\atb{The custom memory controller employs three key sub-modules to facilitate the implementation of new PuM techniques. 
(i) \juan{The \emph{Periodic Operations \revf{Module}}} 
periodically \new{issues} DDR3 refresh~\cite{micron2018ddr3} and interface maintenance commands~\cite{softmc.github}. 
(ii) \juan{A} simple \emph{DDR3 Command Scheduler} 
supports {conventional DRAM operations (e.g., activate, precharge, read, and write)}. \juan{This} scheduler applies an open-bank policy (i.e., DRAM banks are left open following a DRAM row activation) to exploit temporal locality in memory accesses to the DRAM module. {LOAD/STORE memory requests are simply handled by the command scheduler in a latency-optimized way. \new{Thus,} new modules \new{that are implemented to provide new PuM functionality (e.g., a state machine that controls the execution of a new PuM operation)} in the custom memory controller do not compromise the performance of LOAD/STORE memory requests.}}} {(iii) \Copy{R3/6}{\juan{The \emph{Configuration Register File}} (CRF) comprises 16 user-programmable registers that store \newnew{the violated} timing parameters used for DDRx sequences that trigger PuM operations \newnew{(e.g., activation latency used in generating true random numbers using D-RaNGe~\cite{kim.hpca19}, \newnew{see Section~\ref{sec:drange}})} and miscellaneous parameters for PuM implementations (e.g., true random number generation period for D-RaNGe, \newnew{see Section~\ref{sec:drange}}). {In our implementation, CRF stores only the timing parameters used for performing PuM operations (e.g., RowClone and D-RaNGe). We do not store every standard DDRx timing parameter (i.e., non-violated, \omi{which are used exactly as} defined as in DRAM chip specifications) in the CRF. Instead these timings are embedded in the \newnew{command} scheduler.}}}

\begin{figure*}[!t]
  \centering
  \includegraphics[width=0.9\textwidth]{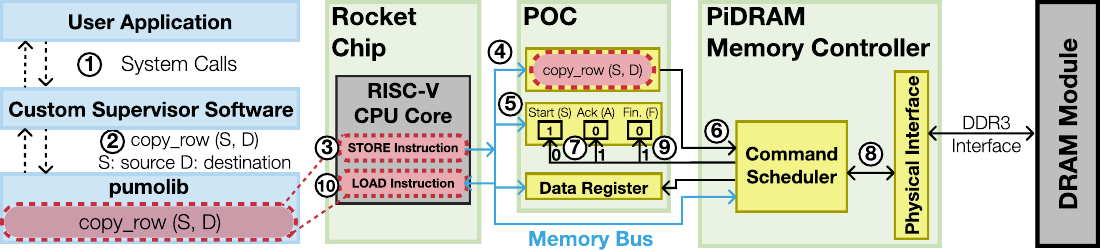}
  \caption{{Workflow for a PiDRAM RowClone-Copy operation}}
  \label{fig:flow}
  
\end{figure*}
\subsection{Software Components}
\label{sec:software-components}
{PiDRAM comprises two key software components that complement {and control} PiDRAM's hardware components {to} provid{e} a flexible and easy to use {end-to-end} PuM framework.} 


\textbf{\ding{184} PuM Operations Library (pumolib).} {The extensible library (\emph{PuM} \emph{o}perations \emph{lib}rary) allows system designers to implement software support for PuM techniques. Pumolib contains customizable functions that interface with POC to perform PuM operations in real unmodified DRAM chips. The customizable functions \newnew{hide} the hardware implementation details of PuM techniques implemented in \X{} \newnew{from software developers (that use pimolib)}.} For example, {although we expose PuM techniques to software via memory LOAD/STORE operations (POC is exposed as a memory-mapped module, Section~\ref{sec:hardware-components}), PuM techniques can also be exposed via specialized instructions provided by ISA extensions.} Pumolib \newnew{hides} such implementation details from the user of the library and \atb{contributes to the modular design of the} framework. 


We implement a general protocol that defines how {programmers} express the information required to execute PuM operations to the PuM operations controller (POC). {A typical function in {pumolib} performs a PuM operation in four steps: \Copy{R3/3}{It (i) writes a PiDRAM instruction to {the} POC's \emph{instruction} register, (ii) sets the \emph{Start} {flag} in POC's \emph{flag} register, (iii) waits for {the} POC to set the \emph{Ack} {flag} in POC's \emph{flag} register, and (iv) reads the result of the PuM operation from POC's \emph{data} register {(e.g., the true random number after performing a{n in-DRAM true random number generation} operation, Section~\ref{sec:drange})}. {We list the currently implemented pumolib functions in Table~\ref{table:pumolib}.}}}


\textbf{\ding{185} Custom Supervisor Software.}\revdel{End-to-end implementation of PuM techniques requires modifications across the hardware and software stack.} \X provides a custom supervisor software that \newnew{implements} the necessary OS primitives (i.e., virtual memory \juan{management}, memory \juan{allocation and alignment) \newnew{for end-to-end implementation of PuM techniques}.} 
{This facilitates developing end-to-end integration of PuM techniques {in the system} as these techniques require modifications across the software stack. For example, integrating RowClone end-to-end {in the full system} requires a new memory allocation mechanism (Section~\ref{sec:rowclone_alignment}) that can satisfy the memory allocation constraints of RowClone~\cite{seshadri2013rowclone}. \new{Thus, we implement the necessary functions and data structures in the custom supervisor software to implement an allocation mechanism that satisfies RowClone's constraints. This allows \X{} to be extended easily to implement support for new PuM techniques that share similar memory allocation constraints (\newnew{e.g., Ambit~\cite{seshadri.micro17}, SIMDRAM~\cite{hajinazarsimdram}, and QUAC-TRNG~\cite{olgun2021quactrngieee}, as shown in} Table~\ref{table:use-cases}).}}


\subsection{Execution of a PuM Operation}
\label{sec:execution-overview}

{We describe the general workflow for a PiDRAM operation (e.g., RowClone-Copy~\cite{seshadri2013rowclone}, random number generation using D-RaNGe~\cite{kim.hpca19}) in Figure~\ref{fig:flow} over an example \texttt{{copy\_row()}} function that is called by the user to {perform a RowClone-Copy operation} in DRAM.}

{{T}he user makes a system call to the custom supervisor software \scalebox{1.1}{{\ding{172}}} that in turn calls the {\texttt{copy\_row(source, destination)}} function in the {pumolib} \scalebox{1.1}{{\ding{173}}}. The function executes {two} store instructions in the RISC-V core \scalebox{1.1}{\ding{174}}. {The first store} instruction {update{s}} the \emph{instruction} register with the {copy\_row} instruction (i.e., the instruction that performs a {RowClone-Copy} operation in DRAM) \scalebox{1.1}{{\ding{175}}} and {the second store instruction} {set{s} the Start flag in the flag register to logic-1 \scalebox{1.1}{\ding{176}} in POC.} {When the Start flag is set,} POC instructs the PiDRAM memory controller to perform a {RowClone-Copy} operation using violated timing parameters \scalebox{1.1}{\ding{177}}. {{T}he {POC waits until the memory controller starts executing the {operation, after which it}} sets the Start flag to logic-0 and the Ack {flag} to logic-1 \scalebox{1.1}{\ding{178}}{, indicating that it started the execution of the PuM operation}.} {T}he PiDRAM memory controller performs the {RowClone-Copy} operation by issuing a set of DRAM commands with violated timing parameters \scalebox{1.1}{{\ding{179}}}. {{When the last DRAM command is issued, the memory controller} sets the Finish flag (denoted as Fin. in Figure~\ref{fig:flow}) in the flag register to logic-1 \scalebox{1.1}{{\ding{180}}}, indicating the end of execution for the last PuM operation that the memory controller acknowledged.} {The copy function periodically checks {either} the Ack {or the} Finish flag in the flag register {(depending on a user-supplied argument)} by executing load instructions that target the flag register \scalebox{1.1}{\ding{181}}. {When the periodically checked flag is set, the copy function returns.} This way, the copy function optionally blocks until the start {(i.e., the Ack flag is set)} or the end {(i.e., the Finish flag is set)} of the execution of the PuM operation (in this example, RowClone-Copy).\footnote{{The data register is not used in \newnew{a} RowClone-Copy~\cite{seshadri2013rowclone} operation because the result of the RowClone-Copy operation is stored {\emph{in memory}} (i.e., the source {memory row} is copied to the destination {memory row}). The data register is used in \newnew{a} D-RaNGe~\cite{kim.hpca19} operation{, as described in Section~\ref{sec:drange}}. {When used, t}he command scheduler store{s} the random numbers generated by {the} D-RaNGe operation in the data register. To read the generated random number, we implement a pumolib function {called} \texttt{rand\_dram()} that executes load instructions in the {RISC-V} core to retrieve the random number from the data register in POC.}}}}


\begin{table*}[b]
    \centering
    \scriptsize
    \caption{\omi{Various} \newnew{known} PuM techniques that can be studied using \X. PuM techniques we implement \juan{in this work} are highlighted in bold.}
    \hspace{1em}
    \begin{tabular}{m{12em}m{10em}m{45em}}
    \toprule
    \textbf{PuM Technique} & \textbf{Description} & \textbf{Integration Challenges} \\
    \midrule
    {\textbf{ComputeDRAM-based~\textbf{\cite{gao2020computedram}}}} \textbf{RowClone~\cite{seshadri2013rowclone}} & Bulk data-copy \juan{and initialization} within DRAM & (i) \emph{memory allocation \juan{and alignment} mechanisms} that map source \& destination operands of a copy operation into same DRAM subarray; (ii) \juan{\emph{memory coherence}, i.e.}, source \revdel{\& destination }operand must be up-to-date in DRAM.\\
    \midrule
    \textbf{D-RaNGe}~\cite{kim.hpca19} & True random number generation using DRAM & (i) periodic generation of true random numbers; (ii) \emph{memory scheduling policies} that minimize the interference caused by random number requests. \\
    \midrule
    {ComputeDRAM-based~\cite{gao2020computedram}} Ambit~\cite{seshadri.micro17} & Bitwise operations in DRAM & (i) \emph{memory allocation \juan{and alignment} mechanisms} that map operands of a bitwise operation into same DRAM subarray; (ii) \juan{\emph{memory coherence}, i.e.}, operands of the bitwise operations must be up-to-date in DRAM. \\
    \midrule
    SIMDRAM~\cite{hajinazarsimdram} & \juan{Arithmetic operations in DRAM} & (i) \emph{memory allocation \juan{and alignment} mechanisms} that map operands of an arithmetic operation into same DRAM subarray; (ii) \juan{\emph{memory coherence}, i.e.}, operands of the arithmetic operations must be up-to-date in DRAM; (iii) \juan{\emph{bit transposition}, i.e., operand bits must be laid out vertically in a single DRAM bitline}. \\
    \midrule
    DL-PUF~\cite{kim.hpca18} & Physical unclonable functions in DRAM & \emph{memory scheduling policies} that minimize the interference caused by generating PUF responses. \\
    \midrule
    \reva{QUAC-TRNG~\cite{olgun2021quactrng} \newnew{and Talukder+~\cite{talukder2019exploiting}}} & \reva{True random number generation using DRAM} & \reva{(i) periodic generation of true random numbers; (ii) \emph{memory scheduling policies} that minimize the interference caused by random number requests; (iii) efficient integration of the SHA-256 cryptographic hash function.} \\
    \bottomrule
    \end{tabular}
    
    \label{table:use-cases}
    
\end{table*}
\subsection{Use Cases}
\label{sec:use-cases}
\atb{\X is primarily designed to study end-to-end implementations of {commodity DRAM based} PuM techniques~\cite{olgun2021quactrng,gao2020computedram,kim.hpca18,kim.hpca19,talukder2019exploiting} on real systems. {Beyond commodity DRAM based PuM techniques}, many prior works propose minor modifications to DRAM arrays to enable various arithmetic~\cite{hajinazarsimdram,deng.dac2018,ferreira2021pluto,angizi2019graphide} and bitwise operations~\cite{seshadri.micro17,seshadri2020indram,seshadri.bookchapter17.arxiv,Seshadri:2015:ANDOR,angizi2019graphide} and security primitives~\cite{orosa2021codic}.}
{These PuM techniques share common memory allocation and coherenc\newnew{e} requirements (Section~\ref{sec:rowclone_alignment}) that must be satisfied to enable their end-to-end integration {into a real system.} \X{} facilitates the implementation of PuM techniques and enables rapid exploration of such integration challenges on a real DRAM-based system.} \newnew{Table ~\ref{table:use-cases} describes some of the \juan{PuM} case studies \X can enable.}

{\new{Other than providing an easy-to-use basis for end-to-end implementations of commodity DRAM based PuM techniques,} \X{} can be \new{easily} extended with a programmable microprocessor placed near the memory controller to study system integration challenges of Processing-near-Memory (PnM) techniques (e.g., efficient pointer chasing~\cite{impica, hashemi.isca16,cont-runahead}, \new{general-purpose compute~\cite{upmem2018}, machine learning~\cite{kwon2021fimdram, ke2021near, kim2021aquabolt, lee2022isscc, niu2022isscc}, \newnew{databases~\cite{lee2022improving,boroumand2019conda,boroumand2016pim}}, \omi{graph processing~\cite{besta2021sisa}}}).\revdel{ \atb{\X} can be deployed on appropriate FPGA boards as a testbed for new memory devices with compute capability~\cite{upmem2018,kwon2021fimdram}.}}

\subsection{{PiDRAM's HW \& SW Components: Summary}}
\label{sec:component-summary}

\Copy{R3/2}{
{We identify and build two hardware components (PuM Operations Controller and Custom Memory Controller) and two software components (PuM Operations Library, Custom Supervisor Software) as key components that \new{are} commonly required by \new{end-to-end PuM implementaions}. We reuse these key components to implement two different PuM mechanisms (RowClone in Section~\ref{sec:rowclone} and D-RaNGe in Section~\ref{sec:drange}) in PiDRAM. The key components can be reused in the same way {to} implement other PuM mechanisms (e.g., the ones in Table~\ref{table:use-cases}). However, reusing a component does not mean that the component can {simply} be instantiated in a system and the system will be able to perform PuM operations immediately.} 

{We acknowledge that these components require modifications to implement new PuM techniques in PiDRAM and possibly to integrate PiDRAM into other systems. In fact, we quantify the degree of these modifications in our RowClone and D-RaNGe case studies. We show that the key components form a useful and easy-to-extend basis for PuM techniques with our Verilog and C code complexity analyses for both use cases (Sections~\ref{sec:rowclone-experimental-methodology} and~\ref{sec:drange-evaluation}).}
}
\subsection{\X Prototype}
\label{sec:prototype}

\begin{figure}[!h]
    \centering
    \includegraphics[width=1.0\linewidth]{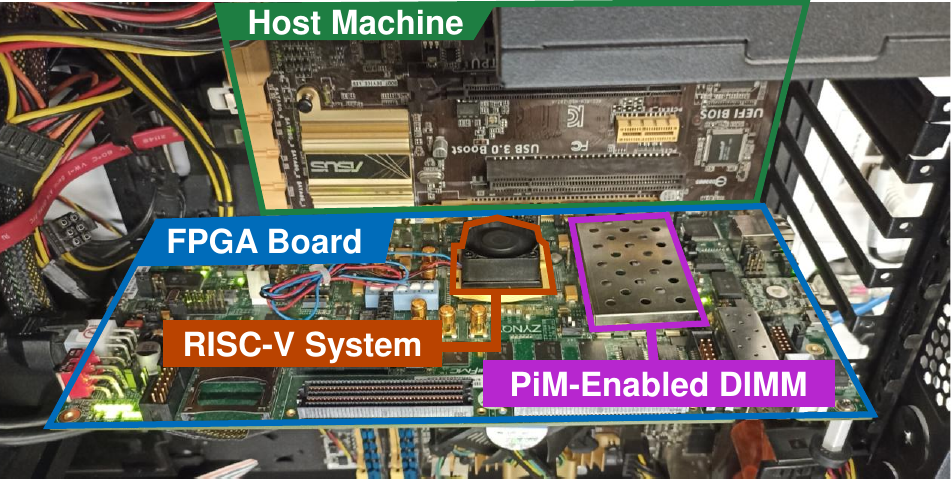}
    \caption{PiDRAM's FPGA prototype}
    \label{fig:prototype}
\end{figure}

\Copy{R3/7B}{\atb{We develop a prototype of the \X framework on an FPGA-based platform. We use the Xilinx ZC706 FPGA board~\cite{zc706} to interface with real DDR3 modules. {Xilinx provides a DDR3 PHY IP~\cite{virtex7mig} that exposes a low-level ``DFI'' interface~\cite{dfi} to the DDR3 module on the board. We use this interface to issue DRAM commmands to the DDR3 module.} We \newnew{use} the existing RISC-V based SoC generator, \omi{Rocket Chip}~\cite{asanovic2016rocket}, to {generate the RISC-V hardware system}. Our custom supervisor software extends the RISC-V \newnew{Proxy Kernel~\cite{riscv-pk}} to support the necessary OS primitives on \X's prototype.} Figure~\ref{fig:prototype} shows our prototype.}

\noindent
\Copy{R1/1B}{{{\textbf{\new{Simulation Infrastructure.}} To \newnew{aid} the users \newnew{in testing}} the correctness of any modifications \newnew{they make} on top of PiDRAM, we provide the developers with a Verilog simulation {environment} that injects regular READ/WRITE commands and custom commands (e.g., update {the Configurable Register File (CRF)}, perform RowClone-Copy, generate random numbers) to the memory controller. When used in conjunction with the Micron DDR3 Verilog model provided by Xilinx~\cite{virtex7mig}, the simulation \newnew{environment} can help the developers \newnew{to} easily understand if something unexpected is happening in their implementation (e.g., {if} timing {parameters} are violated).}}

\noindent
\omi{\textbf{Open Source Repository.} We make PiDRAM freely available to the research community as open source software at \url{https://github.com/CMU-SAFARI/PiDRAM}. Our repository includes the full PiDRAM prototype that has RowClone (Section~\ref{sec:rowclone}) and D-RaNGe (Section~\ref{sec:drange}) implemented end-to-end on the RISC-V system.}

\section{Case Study \#1: End-to-end RowClone}


\label{sec:rowclone}

\revdel{RowClone~\cite{seshadri2013rowclone} proposes \atb{minor changes to DRAM {chips} to copy data in DRAM to mitigate data movement overheads}. 
ComputeDRAM~\cite{gao2020computedram} demonstrates in-DRAM copy operations on contemporary, off-the-shelf DDR3 chips. \atb{Their results show that c}urrent DRAM devices can \atb{reliably} perform copy operations {at different temperatures and supply voltage levels using a set of violated tRAS and tRP timing parameters} in DRAM row granularity. \Copy{R3/1}{{None of the relevant prior works~\cite{seshadri.micro17,seshadri2013rowclone,wang2020figaro,gao2020computedram} provide a clear description {or a real system demonstration (like we do)} of a working memory allocation mechanism that can be integrated into a real operating system to expose RowClone capability to the programmer.}}}

We implement support for ComputeDRAM-\new{based} (i.e., using carefully-engineered sequences of valid DRAM commands {with violated timing parameters}) \new{RowClone (in-DRAM copy/initialization)} operations on PiDRAM to conduct a detailed study \newnew{of}\revdel{ (i) the system performance benefits that RowClone can provide and (ii)} the challenges associated with implementing RowClone end-to-end on a real system. \new{\Copy{R3/1}{{None of the relevant prior works~\cite{seshadri.micro17,seshadri2013rowclone,wang2020figaro,gao2020computedram,seshadri2020indram,seshadri.bookchapter17.arxiv,seshadri.thesis16,hajinazarsimdram} provide a clear description {or a real system demonstration} of a working memory allocation mechanism that can be \new{implemented in} a real operating system to expose RowClone capability to the programmer.}}} \revdel{\new{Using our real system prototype, we study the performance benefits that RowClone can provide in detail.}}

\subsection{Implementation Challenges}

\noindent
\textbf{{Data Mapping.}}
\label{sec:rowclone_alignment}
{RowClone has data mapping and alignment requirements that {cannot be} satisfied by current memory allocation mechanisms (e.g., malloc~\cite{malloc}). We identify four major issues that complicate the process of implementing support for RowClone in real systems. First, \newnew{the} source and destination operands \new{(i.e., \omi{page (4 KiB)-sized} arrays)} of the copy operation must reside in the same DRAM subarray. We refer to this as the \emph{mapping} problem. Second, the source and destination operands must be aligned to DRAM rows. We refer to this as the \emph{alignment} problem. Third, the size of the copied data must be a multiple of the DRAM row size. The size constraint defines the granularity at which we can perform bulk-copy operations using RowClone. We refer to this as the \emph{granularity} problem.} Fourth, \new{RowClone must operate on up-to-date data that reside\newnew{s} in main memory.} Modern systems employ caches to exploit locality in memory accesses and reduce memory latency. \new{Thus,} cache blocks \new{(typically 64 B)} of either the source or the destination operand\newnew{s} of the RowClone operation {may} have \newnew{cache block} copies present in the cache hierarchy. Before performing RowClone, {the} cached copies \newnew{of pieces of both source and destination operands} must be invalidated and written back to main memory\revdel{ if necessary}. We refer to this as the 
{\emph{memory coherence}} problem.

\begin{figure}[h]
  \centering
  \includegraphics[width=\linewidth]{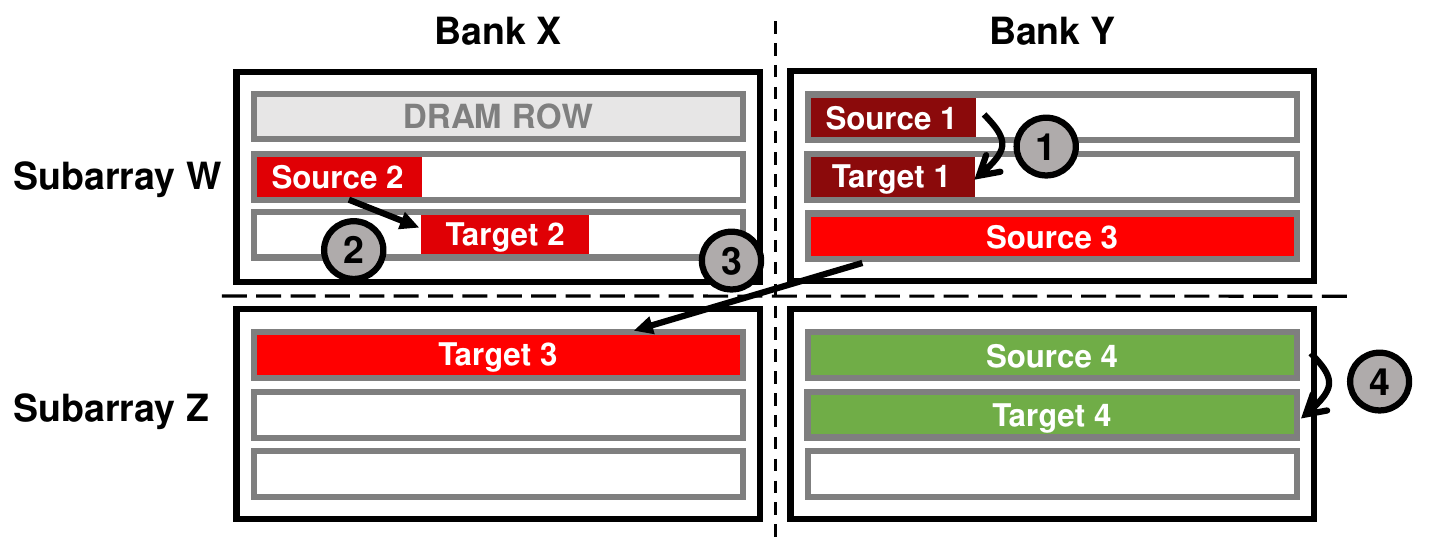}
  \caption{A DRAM {chip} with two banks and two subarrays. Only \omi{one} operation \omi{(i.e., operation~\newnew{\boldfour{}})} can succeed as its operands satisfy \omi{all of} \emph{mapping}, \emph{alignment} and \emph{granularity} constraints.}
  \label{fig:rowclone_alignment}
\end{figure}

\new{We explain the data mapping and alignment requirements of RowClone \newnew{using} Figure~\ref{fig:rowclone_alignment}.} \reve{\new{The figure} depicts a \new{simplified version of a} DRAM {chip} with two banks and two subarrays. The {operand} Source 1 cannot be copied to the {operand} Target 1 as \new{the operands} do not satisfy the \emph{granularity} constraint (\boldone). Performing such a copy operation would overwrite the \newnew{remaining (i.e., non-Target 1)} data in \new{Target 1's DRAM row} with \newnew{the remaining (i.e., non-Source 1)} data in \new{Source 1's DRAM row}. Source 2 cannot be copied to Target 2 as Target 2 is not \emph{aligned} to its DRAM row (\boldtwo). Source 3 cannot be copied to Target 3, as these {operands} are not \emph{mapped} to the same DRAM subarray (\boldthree). \new{In contrast}, Source 4 can be copied to Target 4 using in-DRAM copy \new{because these operands} are (i) \emph{mapped} to the same DRAM subarray, (ii) aligned to their DRAM rows and (iii) occupy their rows completely (i.e., the {operands} have sizes equal to DRAM row size) (\boldfour).}

\subsection{Memory Allocation Mechanism}
\label{sec:rowclone_mechanism}

\new{C}omputing systems employ various layers of address mappings that obfuscate the DRAM row-bank-column address mapping from the programmer~\cite{helm2020Reliable,cojocar2020susceptible}, \new{which makes allocating source and target operands as depicted in Figure~\ref{fig:rowclone_alignment}-(\boldfour) difficult}. \new{DRAM manufacturers employ DRAM\newnew{-}internal address mapping schemes (Section~\ref{sec:background-dram}) that translate from logical (e.g., \newnew{memory-controller-visible} DRAM row, bank, column) \omi{addresses} to physical DRAM addresses.} {\new{G}eneral-purpose} processors use complex functions to map physical addresses to DDRx addresses \new{(e.g., DRAM banks, rows and columns)}~\cite{hillenbrand2017Physical}. \new{The} operating system (OS) maps virtual addresses to physical addresses to provide isolation between multiple processes\revdel{ and 
\juan{the} illusion of larger-than-available physical memory}. \new{Only these virtual addresses are exposed to the programmer}. Without control over the virtual address $\rightarrow{}$ DRAM address mapping, the programmer \emph{cannot} \new{easily} place data in a way 
\juan{that satisfies} the mapping and alignment requirements of \newnew{an} in-DRAM copy operation. 



\begin{figure}[!ht]
  \centering
  \includegraphics[width=\linewidth]{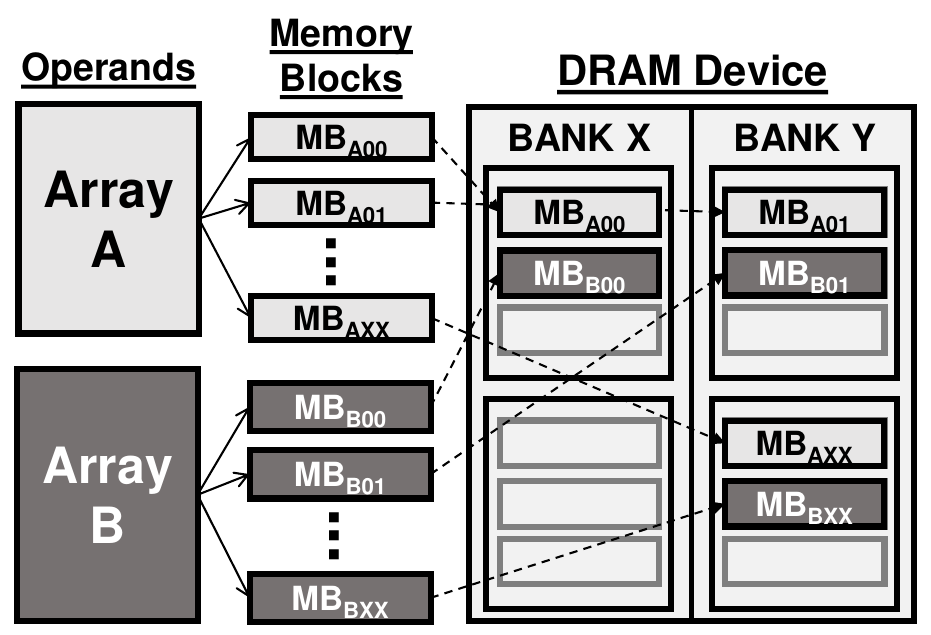}
  \caption{Overview of our memory allocation mechanism\revdel{ Source and destination operands (Arrays A and B) \newnew{that are} comprised of page-sized memory blocks are placed in DRAM subarrays.}}
  \label{fig:memory-allocation-mechanism}
\end{figure}

\new{W}e implement a new memory allocation mechanism that {can perform} memory allocation for \new{RowClone (in-DRAM copy/initialization)} operations. \atb{\new{This} mechanism enables page-granularity \new{RowClone} operations (i.e., a virtual page can be copied to another virtual page using RowClone) \emph{without} introducing any changes to the programming model.} \new{The mechanism} places the operands of RowClone operations \new{in} the same DRAM subarray while maximizing the bank-level parallelism in regular DRAM accesses (reads \& writes) to these operands \new{(such that the commonly\newnew{-}performed streaming accesses to these operands benefit from bank-level parallelism in DRAM)}. {\atb{Figure~\ref{fig:memory-allocation-mechanism} depicts an overview of our memory allocation mechanism.}}

\new{At a high level, o}ur \new{memory allocation} mechanism \new{(i)} splits the source and destination operands into page-sized virtually-addressed memory blocks, \new{(ii)} allocate\new{s} two physical pages in different DRAM rows in the same DRAM subarray,\revdel{ These physical pages are aligned to the same page-size boundary (i.e., satisfying the \emph{alignment} requirement).} \new{(iii) assigns} these physical pages to virtual pages that correspond to the source and destination memory blocks at the same index such that the source block can be copied to the destination block using RowClone. We repeat this process until we exhaust the page-sized memory blocks. \new{As the mechanism processes subsequent page-sized memory blocks of the \newnew{two} operands, it allocates} physical pages from a different DRAM bank \new{to maximize bank-level parallelism in streaming accesses to these operands}.

\begin{figure*}[!ht]
  \centering
  \includegraphics[width=.8\textwidth]{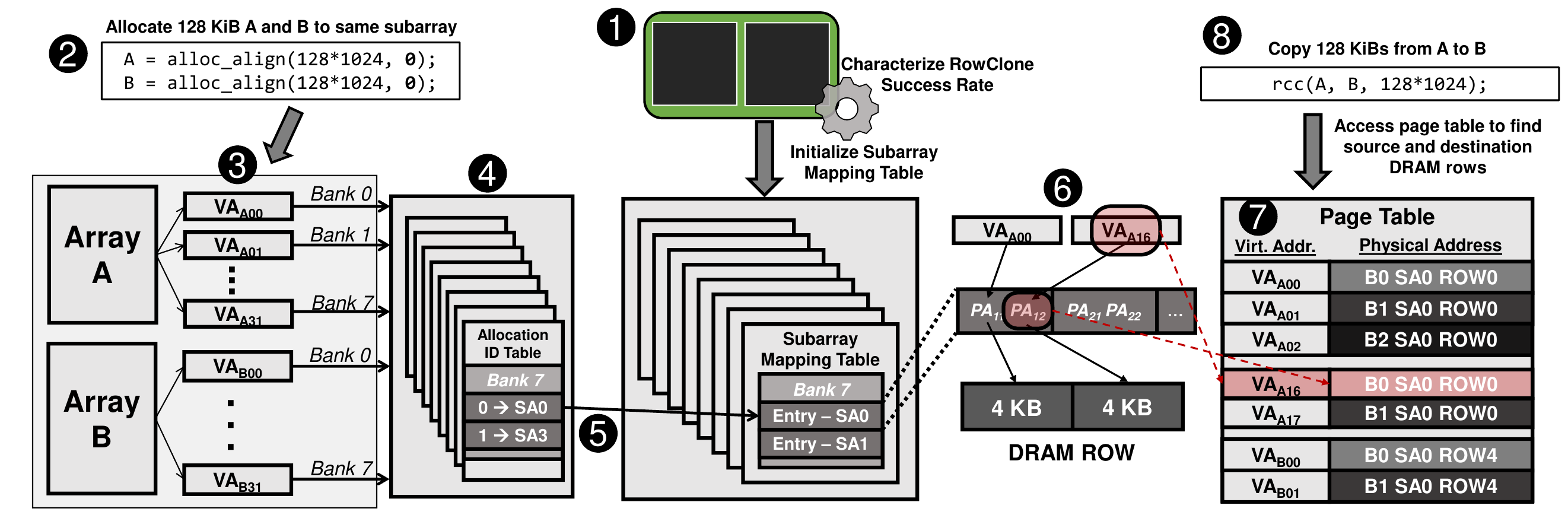}
  \caption{{\texttt{Alloc\_align() and RowClone-Copy (\texttt{rcc}, see Section~\ref{sec:rccrci}) workflow.}}} 
  \label{fig:alloc-align-walkthrough}
  
\end{figure*}

To overcome the \emph{mapping}, \emph{alignment}, and \emph{granularity} problems, we implement our memory management mechanism in the custom supervisor software of \X.
We expose \new{the allocation mechanism} \newnew{using} the \texttt{alloc\_align(N, ID)} system call. The system call returns a pointer to a contiguous array of \emph{N} bytes in the virtual address space {(i.e., one operand)}. \Copy{R5/5}{Multiple calls with the same \emph{ID} to \texttt{alloc\_align(N, ID)} place the allocated arrays in the same subarray in DRAM, such that they can be copied from one to another using RowClone. {If \emph{N} is too large such that it exceeds the size of available physical memory, \texttt{alloc\_align} fails and causes an exception.} \revdel{Figure~\ref{fig:alloc-align-walkthrough}\revdel{shows the components of our mechanism and} presents the workflow of \texttt{alloc\_align()}.}} \revd{Our implementation of\revdel{\texttt{alloc\_align}} \new{RowClone} requires application developers to directly use \texttt{alloc\_align} to allocate data instead of \texttt{malloc} and similar function calls.}

{\new{The \omi{custom supervisor software}} maintains three key structures \new{to make \texttt{alloc\_align()} work}: (i) Subarray Mapping Table (SAMT), (ii) Allocation ID Table (AIT), and (iii) Initializer Rows Table (IRT).}

\noindent
\textbf{{1) Subarray Mapping Table (SAMT).}} We use the \textbf{S}ubarray \textbf{Ma}pping \textbf{T}able (SAMT) to maintain a list of physical page addresses that point to DRAM rows that are in the same DRAM subarray. {\new{\texttt{alloc\_align()}} queries SAMT to find physical addresses that map to rows in one subarray.} 

\new{SAMT} contains the physical pages that point to DRAM rows in each \new{subarray}. \new{SAMT is} indexed using subarray identifiers (SA IDs) in the range \emph{[0, number of subarrays\revdel{ in a DRAM bank})}. \new{\new{SAMT}} contains an entry for every subarray\revdel{ in the DRAM \new{bank}}. \new{An} entry consists of two elements: (i) the number of free physical address tuples and (ii) a list of physical address tuples. Each tuple in the list contains two physical addresses that \newnew{respectively} point to the first and second hal\newnew{ves} of the same DRAM row. The list of tuples \newnew{contains} all the physical addresses that point to DRAM rows in the DRAM subarray indexed by the SAMT\revdel{ \new{sub-table}} entry. We allocate free physical pages \new{listed in an} entry and assign them to the virtual pages (i.e., memory blocks) that make up the {row-copy operands (i.e., arrays)} allocated by \texttt{alloc\_align()}. We slightly modify our \new{high-level memory} allocation mechanism to allow for two memory blocks \new{(4 KiB virtually-addressed pages)} of an array to be placed in the same DRAM row, as the page size in our system is 4 KiB, and the size of a DRAM row is 8 KiB. \atb{We call two memory blocks in {the same operand} that are placed in the same DRAM row \emph{sibling memory blocks} \omi{(also called sibling pages)}. The parameter \emph{N} \new{of} the \texttt{alloc\_align()} call defines this relationship: We designate memory blocks that are precisely {\emph{N/2}} bytes apart as \emph{sibling memory blocks}.}

\noindent
\textbf{{Finding \new{the} DRAM Rows in a Subarray.}} \Copy{R2/1A}{{Finding the DRAM row addresses that belong \newnew{to} the same subarray is not straightforward due to DRAM-internal mapping schemes employed by DRAM manufacturers (Section~\ref{sec:rowclone_alignment}). It is extremely difficult to learn which DRAM address (i.e., bank-row-column) is actually mapped to a physical location (e.g., a subarray) in the DRAM device, as these mappings are not exposed through publicly accessible datasheets or standard definitions~\cite{jedecDDR4,micron2016ddr4,patel2022case}. We make the key observation that the entire mapping scheme need \emph{not} be available to successfully perform RowClone operations.}}

\Copy{R2/1B}{We observe that for a set of \emph{\{source, destination\}} DRAM row address pairs, RowClone operations \atb{repeatedly} succeed with a 100\% probability. We hypothesize that these pairs of DRAM row addresses are mapped to the same DRAM subarray. {We identify these row address pairs by conducting a \emph{\newnew{RowClone success rate}} experiment where we repeatedly perform RowClone operations between every \emph{source, destination} row address pair in a DRAM bank. Our experiment works in three steps{:} we (i) initialize both the source and the destination row with random data, (ii) perform a RowClone operation from the source to the destination row, and (iii) compare the data in the destination row with the source row. \omi{RowClone success rate is calculated as the number of bits that differ between the source and destination rows' data divided by the number of bits stored in a row (8 KiB in our prototype).} If there is no difference between the source and the destination rows' data \omi{(i.e., the RowClone success rate for the source and the destination row is 100\%)}, we {infer that} the RowClone operation {was} successful. We repeat the experiment for 1000 iterations for each row address pair and if every iteration is successful, we store the address pair in the SAMT{, indicating that the row address pair is mapped to different rows in the same DRAM subarray}.}} \Copy{R4/9}{{\newnew{The same RowClone success rate experiment} could be conducted \revdel{to identify DRAM row address pairs {that are mapped to the same DRAM subarray (i.e., data can be copied from one row address to the other row address in the pair} using RowClone operations) }in \newnew{other} systems that are based on PiDRAM or in a PiDRAM prototype that uses a different DRAM {module}. Since \newnew{the RowClone success rate experiment} is a one-time process, its overheads (e.g., time taken to iterate over all DRAM rows using our experiment) \omi{are} amortized over the lifetime of such a system.}}

\noindent
\textbf{{2) Allocation ID Table (AIT).}}
To {keep track of different operands} that are allocated by \texttt{alloc\_align} using the same \emph{ID} \newnew{(used to place different arrays in the same subarray)}, we use the \textbf{A}llocation \textbf{I}D \textbf{T}able (AIT).\revdel{ AIT is logically partitioned into sub-tables for each DRAM bank.} AIT entries are indexed by \emph{allocation ID}s (the parameter \emph{ID} \new{of} the \emph{alloc\_align} call). Each AIT entry stores a pointer to an SAMT entry. The SAMT entry pointed by the AIT entry contains the set of physical addresses that were allocated using the same \emph{allocation ID}. AIT entries are used by the \texttt{alloc\_align} function to find which DRAM subarray can be used to allocate DRAM rows from, such that the newly allocated array can be copied to other arrays allocated using the same \emph{ID}.

\noindent
\textbf{{3) Initializer Rows Table (IRT).}}
To find which row in a DRAM subarray can be used as the source operand in zero-initialization (RowClone-Initialize) operations, we maintain the \textbf{I}nitializer \textbf{R}ows \textbf{T}able (IRT). The IRT is indexed using physical page numbers. RowCopy-Initialize operations query the IRT to obtain the physical address of the DRAM row that is initialized with zeros and \new{that belong} to the same subarray as the destination operand (i.e., the DRAM row to be initialized with zeros).

\atb{Figure~\ref{fig:alloc-align-walkthrough} describes how \texttt{alloc\_align()} works over an end-to-end example. \newnew{Using the RowClone success rate experiment (\omi{described above}), the custom supervisor software (\omi{CSS for short})} find\newnew{s} the DRAM rows that are in the same subarray (\boldone) and initialize\newnew{s} the \newnew{Subarray Mapping Table (SAMT)}. The programmer allocates two 128 KiB arrays, A and B, via \texttt{alloc\_align()} using the same \emph{allocation id} (\textbf{0}), with the intent to copy from A to B (\boldtwo). \newnew{\omi{CSS}} allocate\newnew{s} contiguous ranges of virtual addresses to A and B, \newnew{and} then split\newnew{s the virtual address ranges} into page-sized memory blocks (\boldthree). \newnew{\omi{CSS}} assigns consecutive memory blocks to consecutive DRAM banks and access\new{es} the \newnew{Allocation ID Table (AIT)} with the \emph{allocation id} (\boldfour) for each memory block. \newnew{By accessing the AIT, \omi{CSS} retrieves} the \emph{subarray id} that points to a SAMT entry. The SAMT entry corresponds to the subarray that contains the arrays that are allocated using the \emph{allocation id} (\boldfive). \newnew{\omi{CSS}} access\newnew{es} the SAMT entry to retrieve two physical addresses that point to the same DRAM row. \newnew{\omi{CSS}} map\newnew{s} a memory block and its \emph{sibling \omi{memory block}} \new{(i.e., the memory block that is N/2 bytes away from this memory block, where N is the \emph{size} argument of the \texttt{alloc\_align()} call)} to these two physical addresses, such that they are mapped to the first and the second halves of {the same} DRAM row (\boldsix). \reva{Once allocated, these physical addresses are pinned to main memory and cannot be swapped out to storage.} Finally, \newnew{\omi{CSS}} update\newnew{s} the page table with the physical addresses to map the memory blocks to the same DRAM row (\boldseven).}

\subsection{Maintaining \juan{Memory Coherence}}

\label{sec:coherency}
{Since memory instructions update the cached copies of data \newnew{(Section~\ref{sec:rowclone_alignment})},} a naive implementation of RowClone can potentially operate on stale data because cached copies of RowClone operands can be modified \new{by CPU store instructions}. 
\juan{\new{Thus,} we need to ensure memory coherence to prevent RowClone from operating on stale data.}

\new{W}e implement a new \juan{custom RISC-V} instruction, \juan{called \emph{CLFLUSH},} to flush dirty cache blocks to DRAM 
(RISC-V does not implement any cache management operations~\cite{riscv-spec}) \juan{so as to ensure RowClone operates on up-to-date data}. 
\juan{\newnew{A} CLFLUSH \newnew{instruction} flushes \newnew{(invalidates)}} \newnew{a} {physically addressed} dirty \newnew{(clean)} cache block. \atb{CLFLUSH or other cache management operations with similar semantics are supported in X86~\cite{x86-manual} and ARM architectures~\cite{arm-cmos}. \new{Thus, the CLFLUSH instruction (that we implement) provides a minimally invasive solution (i.e., it requires no changes to the specification of commercial ISAs) to the memory coherence problem.}}

\revdel{ \new{This way, the effects of CLFLUSH on the performance of end-to-end RowClone (Section~\ref{sec:rowclone-experimental-methodology}) \revdel{likely}represents \new{a case} of a minimally invasive solution (i.e., it requires no changes to the specification of commercial ISAs).}} 
We modify the non-blocking data cache and the \omi{R}ocket core modules (defined in \textit{NBDCache.scala} and \textit{rocket.scala} in \omi{Rocket Chip}~\cite{asanovic2016rocket}, respectively) to implement CLFLUSH. We modify the RISC-V GNU compiler toolchain~\cite{riscv-gnu-toolchain} to expose CLFLUSH as an instruction to C/C++ applications. 
\juan{Before executing a RowClone \newnew{Copy or Initialization} operation \newnew{(see Section~\ref{sec:rccrci})}, \newnew{the \omi{custom supervisor software}} \new{flushes} \atb{(invalidate\new{s})} the cache blocks \new{of} the source \atb{(destination)} row of the RowClone operation \atb{using CLFLUSH}.}

\subsection{RowClone-Copy and RowClone-Initialize}
\label{sec:rccrci}
\new{W}e support the RowClone-Copy and RowClone-Initialize operations in our custom supervisor software via two functions: (i) \newnew{RowClone-Copy,} \texttt{rcc(void *dest, void *src, int size)} and (ii) \newnew{RowClone-Initialize,} \texttt{rci(void* dest, int size)}. \atb{\texttt{rcc} copies \emph{size} \newnew{number of} contiguous bytes in the virtual address space starting from the \emph{src} memory address to the \emph{dest} memory address. \texttt{rci} initializes \emph{size} \newnew{number of} contiguous bytes in the virtual address space starting from the \emph{dest} memory address.} We expose \texttt{rcc} and \texttt{rci} to user-level programs \newnew{using} system calls \newnew{defined in the \omi{custom supervisor software}}. 

\juan{\texttt{rcc}} \new{(i)} splits the source and destination operands into page-aligned, page-sized blocks, \new{(ii)} traverses the page table \atb{(Figure~\ref{fig:alloc-align-walkthrough}-\boldeight)} to find the physical address of each block (i.e., the address of a DRAM row), \new{(iii)} flushes all cache blocks corresponding to the source \newnew{operand} and \atb{invalidates all cache blocks corresponding to the} destination \newnew{operand}, \new{and (iv)}
performs a RowClone operation from the source row to the destination row using pumolib{'s \texttt{copy\_row()} function}.

\juan{\texttt{rci}} \new{(i)} splits the destination operand into page-aligned, page-sized blocks, \new{(ii)} traverses the page table to find the physical address of \newnew{the destination operand}\new{, (iii)} queries the \newnew{Initializer Rows Table (IRT, see Section~\ref{sec:rowclone_mechanism})} to obtain the physical address of the initializer row \atb{(i.e., source operand)}, \new{(iv)} \atb{invalidates} the cache blocks corresponding to the \newnew{destination operand}, and \new{(v)} performs a RowClone operation from the initializer row to the destination row using using pumolib{'s \texttt{copy\_row()} function}. 

\subsection{Evaluation}

We evaluate our solutions for the challenges in implementing RowClone end-to-end on a real system using \X. We modify the custom memory controller to implement DRAM command sequences \atb{($ACT\rightarrow{}PRE\rightarrow{}ACT$)} to trigger RowClone operations. \atb{We set the $tRAS$ and $tRP$ parameters to 10 $ns$ {(below the manufacturer-recommended \SI{37.5}{\nano\second} for tRAS and \SI{13.5}{\nano\second} for tRP~\cite{micron2018ddr3})}.} We modify our custom supervisor software to implement our memory allocation mechanism and add support for RowClone-Copy \new{(\texttt{rcc})} and RowClone-Initialize \new{(\texttt{rci})} operations.


\subsubsection{Experimental Methodology}
\label{sec:rowclone-experimental-methodology}
\revdel{We use \X to implement RowClone end-to-end \juan{and evaluate our solutions to RowClone's system integration challenges, as explained in Section~\ref{sec:rowclone_alignment}}.} 
Table~\ref{table:system-configuration} describes the configuration of the components in our system. \atb{We use the pipelined and in-order \newnew{R}ocket core with 16 K\newnew{i}B L1 data cache and 4\newnew{-}entry TLB as the main processor of our system. We use the 1 GiB DDR3 module available on the ZC706 board as the main memory where we conduct PuM operations.} 

\begin{table}[!ht]
  \centering
  \caption{PiDRAM system configuration}
  
  \scriptsize
  \begin{tabular}{@{} l @{}}
  \toprule
  \textbf{CPU:} 50~MHz; in-order Rocket core \cite{asanovic2016rocket}; \textbf{TLB} 4 entries DTLB; LRU policy\\        
  \midrule
  \textbf{L1 Data Cache:} 16~KiB, 4-way; 
  64~B line; random replacement policy\\
  \midrule
  \textbf{DRAM Memory:} 1~GiB DDR3; 800MT/s; single rank; 8~KiB row size\\
  \midrule
  \end{tabular}
  
  \label{table:system-configuration}
\end{table}

\atb{Implementing RowClone require\newnew{s} an additional 198 lines of Verilog code over \X's existing Verilog design. We add 43 and 522 lines of C code to pumolib and to our custom supervisor software, respectively, to implement RowClone in the software components.}

\begin{figure}[!ht]
  \centering
  \includegraphics[width=0.40\textwidth]{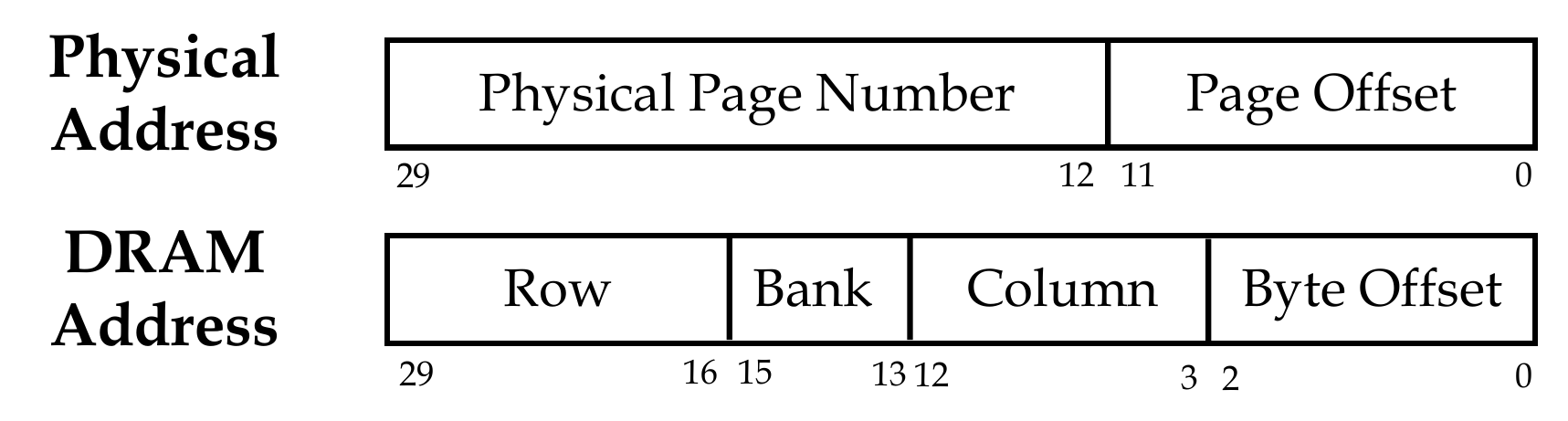}
  \caption{Physical address to DRAM address mapping in \X. \newnew{B}yte offset is used to address the byte in the DRAM burst.}
  \label{fig:DDR-address-mapping}
\end{figure}

Table~\ref{fig:DDR-address-mapping} describes the mapping scheme we use in our custom memory controller to translate from physical to DRAM row-bank-column addresses. We map physical addresses to DRAM columns, banks, and rows from lower-order bits to higher-order bits to exploit the bank-level parallelism in memory accesses to consecutive physical pages. We note that our memory management mechanism is compatible with other physical address $\rightarrow{}$ DRAM address mappings~\cite{hillenbrand2017Physical}. \Copy{R2/2}{{For example, for a mapping {scheme} where \omu{page offset bits (physical address (PA) [11:0]) include all or a subset of the bank address bits}\revdel{\omi{DRAM bank and DRAM row addresses, in that order, are mapped from the most significant to the least significant bits in physical addresses (i.e., physical address (PA) [29:27] maps to DRAM bank address and PA[26:13] maps to DRAM row address)}},\revdel{ bank address and the column address \newnew{in the address scheme depicted in Table~\ref{table:system-configuration}} {are} swapped\revdel{(Table~\ref{fig:DDR-address-mapping}},} a single RowClone operand \omu{(i.e., a 4 KiB page)} would be split across multiple DRAM banks. This \new{only} coarsens the granularity of RowClone operations as the \newnew{{sibling \omi{pages}}} that must be copied in unison, to satisfy the granularity constraint, increases.}} We expect that for \new{other complex or unknown} physical address $\rightarrow{}$ DRAM address mapping scheme\new{s}, the characterization of the DRAM device for RowClone success rate would take longer. In the worst case, DRAM row addresses that belong to the same DRAM subarray can be found by testing all combinations of physical addresses \newnew{for their RowClone success rate}. 

We evaluate \texttt{rcc} and \texttt{rci} operations under two configurations to understand the copy/initialization throughput improvements provided by \texttt{rcc} and \texttt{rci} \newnew{over traditional CPU-copy operations performed by the Rocket core}, and to understand the overheads introduced by end-to-end support for {commodity DRAM based} PuM operations. We test two configurations: (i) \emph{Bare-Metal}, to find the maximum RowClone throughput our implementation provides solely using pumolib, 
and (ii) \emph{No Flush}, to understand the benefits our end-to-end implementation \new{(i.e., with system support)} of RowClone can provide in copy/initialization throughput when data in DRAM is up-to-date \omi{(i.e., when no coherence operations are needed)}.

\noindent
\textbf{Bare-Metal.} We assume that RowClone operations always target data that is allocated correctly in DRAM (i.e., \newnew{there is} no overhead introduced by address translation, IRT accesses, and CLFLUSH operations). We directly issue RowClone operations via pumolib using physical addresses. \newnew{Traditional CPU-copy operations (executed on the Rocket core) also use physical addresses}.

\noindent
\textbf{No Flush.} We assume that the programmer uses the \texttt{alloc\_align} function to allocate the operands of RowClone operations. We use 
{a} version of \texttt{rcc} and \texttt{rci} system calls that do not use CLFLUSH to flush cache blocks of source and destination operands of RowClone operations. We run the \emph{No Flush} configuration on our custom supervisor software. \newnew{Both \texttt{rcc} and \texttt{rci}, and traditional CPU-copy operations use virtual addresses}.


\revcommon{\subsubsection{Workloads}}
\label{sec:methodology-workloads}
For the two configurations, we run a microbenchmark that consists of two programs, \emph{\newnew{copy}} and \emph{\newnew{init}}, on \new{our prototype}. Both programs take the argument $N$, where \emph{copy} copies an $N$-byte array to another $N$-byte array and \emph{init} initializes an $N$\newnew{-}byte array \newnew{to all} zeros. Both programs \atb{have two versions: (i) CPU-copy, which} copies/initializes data using memory loads and stores, (ii) RowClone, which uses RowClone operations to perform copy/initialization. \atb{\new{All programs use} \texttt{alloc\_align} \new{to allocate data}.}
\atb{The performance results we present in this section are the average of a \omi{1000} runs.}
\atb{To maintain the same initial system state for both CPU-copy and RowClone, \revdel{(i.e., initially the data is up-to-date in DRAM), }we flush all cache blocks \newnew{before} each \newnew{one} \new{of the \omi{1000}} runs.}
We run each program for array sizes (\emph{N}) that are powers of two and {$8~KiB$} $< N < 8~MiB$,  
and find the average copy/initialization throughput \newnew{across all 1000 runs} (by measuring the \# of elapsed CPU cycles to execute copy/initialization operations) for CPU-copy, RowClone-Copy (\texttt{rcc}), and RowClone-Initialize (\texttt{rci}).\footnote{\Copy{R1/2}{{We tested RowClone operations using \texttt{alloc\_align()} with up to 8 MiB of allocation size {since} we observed diminishing returns on performance improvement provided by RowClone operations \newnew{on} {larger} array sizes.}}}

\atb{We analyze the overheads of CLFLUSH operations on copy/initialization throughput that \texttt{rcc} and \texttt{rci} can provide. We measure the execution time of CLFLUSH operations \new{in} our \new{prototype} to find how many CPU cycles it takes to flush a (i) dirty and (ii) clean cache block on average \newnew{across 1000 measurements}. We simulate various scenarios \new{(described in Figure~\ref{fig:system-flush-overhead})} where we assume a certain \newnew{fraction} of the operands of RowClone operations are cached and dirty.}


\subsubsection{Bare-Metal RowClone}

\begin{figure}[h] 
  \centering
  \includegraphics[width=\linewidth]{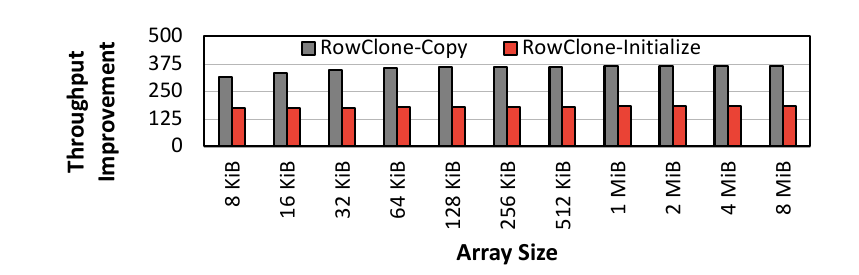}
  \caption{\omi{RowClone-Copy} and \omi{RowClone-Initialize} over traditional CPU-copy and -initialization for the Bare-Metal configuration
  }
  \label{fig:bare-metal-speedup}
\end{figure}

Figure~\ref{fig:bare-metal-speedup} \juan{shows} the throughput improvement \omi{provided by} \texttt{rcc} {and \texttt{rci}} for \emph{copy} {and \emph{initialize}} \omi{over CPU-copy and CPU-initialization}  
for increasing array sizes. 

We make two major observations. \newnew{First, we observe that \texttt{rcc} and \texttt{rci} provide significant throughput improvement over traditional CPU-copy and \omi{CPU}-initialization. The throughput improvement \omi{provided by \texttt{rcc}} ranges from 317.5$\times{}$ (for 8 KiB arrays) to 364.8$\times{}$ (for 8 MiB arrays). \omi{The throughput improvement provided by \texttt{rci} ranges} from 172.4$\times{}$ to 182.4$\times{}$.} \newnew{Second,} the throughput improvement provided by \texttt{rcc} and \texttt{rci} increases as the array size increases. This increase saturates when the array size \newnew{reaches} 1 MiB. 
\omi{The load/store instructions used by CPU-copy and CPU-initialization access the operands in a streaming manner. The eviction of dirty cache blocks (i.e., the destination operands of copy and initialization operations) interfere with other memory requests on the memory bus.\footnote{\omi{Because the data cache in our prototype employs random replacement policy, as the array size increases, the fraction of cache evictions among all memory requests also increase\omi{s}, causing \omu{larger} interference on the memory bus (i.e., more memory requests to satisfy all cache evictions). The interference saturates at 1 MiB array size.}} We attribute the observed saturation at 1 MiB array size to the interference on the memory bus.}

\revdel{Second, the \emph{latency} of a RowClone-Copy operation to copy 8 KiB in DRAM (using pumolib) is only 58 CPU cycles.
\atb{This time is spent on (i) running the pumolib function that executes memory requests (e.g., \omi{\emph{store}} instructions
) to store data in the instruction and flag registers in the POC to ask the memory controller to perform a RowClone operation, (ii) waiting for the memory controller to respond with an acknowledgment}. The latency of executing RowClone-Copy increases \emph{linearly} with the array size. {We make similar observations for RowClone-Initialize. RowClone-Initialize can provide nearly the half of the throughput improvement provided RowClone-Copy, 182.4$\times{}$ over the CPU-Initialization baseline.  Because the CPU needs to execute only half as many instructions compared to copy operations (one load and one store) in initialization operations (one store), it can perform an initialization operation approximately two times faster than a copy operation for the same array size.}}


\subsubsection{\juan{No Flush} RowClone}

\begin{figure}[h] 
  \centering
  \includegraphics[width=\linewidth]{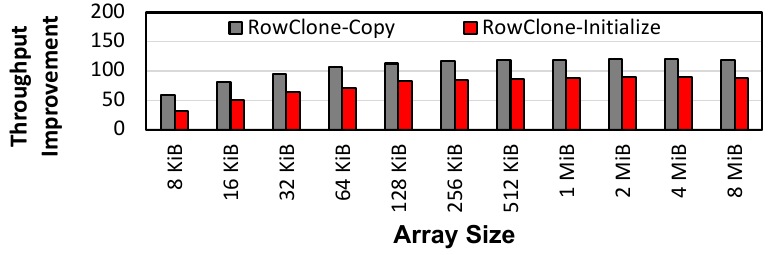}
  \caption{Throughput improvement provided by \newnew{\omi{RowClone-Copy} and \omi{RowClone-Initialize} over traditional CPU-copy and -initialization for the NoFlush configuration}.}
  \label{fig:system-copy-speedup1}
\end{figure}

We analyze the overhead in copy/initialization throughput introduced by system support (Section~\ref{sec:rowclone_mechanism})\revdel{that we implement to enable RowClone end-to-end}.
\reva{Figure~\ref{fig:system-copy-speedup1}} shows the throughput improvement of copy and initialization provided in the \emph{No Flush} configuration by \texttt{rcc} and \texttt{rci} operations.

\revdel{\newnew{Figure~\ref{fig:system-copy-speedup2-abs} shows the execution time of \texttt{rcc} and \texttt{rci} operations in CPU cycles.} Figure~\ref{fig:system-copy-speedup2} shows the proportional increase in \texttt{rcc} and \texttt{rci}'s execution time between consecutive array sizes \newnew{(i.e., the y-axis value for an array size S (between 16 KiB and 8 MiB) shows the execution time of \texttt{rcc} or \texttt{rci} normalized to the execution time of \texttt{rcc} or \texttt{rci} for an array size of S/2)}. For example, the \newnew{circles} at 16 KiB \omi{(2 MiB)} shows the execution time of \texttt{rcc} \newnew{\texttt{and rci}} for 16 KiB \omi{(2 MiB)} arrays divided by the execution time of \texttt{rcc} \newnew{\texttt{and rci}} for 8 KiB \omi{(1 MiB)} arrays. 

\begin{figure}[ht]
     \centering
     \vspace{-2mm}
         \begin{subfigure}[b]{0.47\textwidth}
         \centering
         \includegraphics[width=\textwidth]{figures/perf-absolute.pdf}
         \caption{{The execution time of \texttt{rcc} and \texttt{rci}}}
         \label{fig:system-copy-speedup2-abs}
     \end{subfigure}
     \hfill
     \begin{subfigure}[b]{0.47\textwidth}
         \centering
         \includegraphics[width=\textwidth]{figures/system-copy-init-speedup2.pdf}
         \caption{The increase in \texttt{rcc} and \texttt{rci}'s execution time}
         \label{fig:system-copy-speedup2}
     \end{subfigure}

     \vspace{-3mm}
     \caption{\texttt{rcc} and \texttt{rci} execution time (left) and increase in the execution time as array size increases (right)}
     \vspace{-5mm}
\end{figure}
}

We make two major observations: 
First, \juan{\texttt{rcc}} improves the copy throughput by 58.3$\times{}$ for 8 KiB and by 118.5$\times{}$ for 8 MiB arrays, whereas \texttt{rci} improves initialization throughput by 31.4$\times{}$ for 8 KiB and by 88.7$\times{}$ for 8 MiB arrays.
\atb{Second, we observe that the throughput improvement provided by \texttt{rcc} and \texttt{rci} improves \emph{non-linearly} as the array size increases. The \newnew{\omu{execution time (in Rocket core clock cycles)} of} \texttt{rcc} and \texttt{rci} operations \omu{(not shown in Figure~\ref{fig:system-copy-speedup1})} \emph{does not increase linearly} with the array size. \omu{For example, the execution time of \texttt{rcc} is 397 and 584 cycles at 8 KiB and 16 KiB array sizes, respectively, resulting in a $1.47\times{}$ increase in execution time between 8 KiB and 16 KiB array sizes. However, the execution time of \texttt{rcc} is 92,656 and 187,335 cycles at 4 MiB and 8 MiB array sizes, respectively, resulting in a $2.02\times{}$ increase in execution time between 4 MiB and 8 MiB array sizes. We make similar observations on the execution time of \texttt{rci}.} For every RowClone operation, \texttt{rcc} and \texttt{rci} walk the page table to find the physical addresses corresponding to the source (\texttt{rcc}) and the destination (\texttt{rcc} and \texttt{rci}) operands. We attribute the non-linear increase in \texttt{rcc} and \texttt{rci}'s execution time to (i) the locality exploited by \newnew{the R}ocket core in accesses to the page table and (ii) the \revdel{proportionally }diminishing constant cost in \newnew{the} execution time \newnew{of both \texttt{rcc} and \texttt{rci}} due to common instructions executed to perform a system call.}


\subsubsection{CLFLUSH Overhead}
\label{sec:clflush-overhead}
\begin{figure}[!h] 
  \centering
  \includegraphics[width=.47\textwidth]{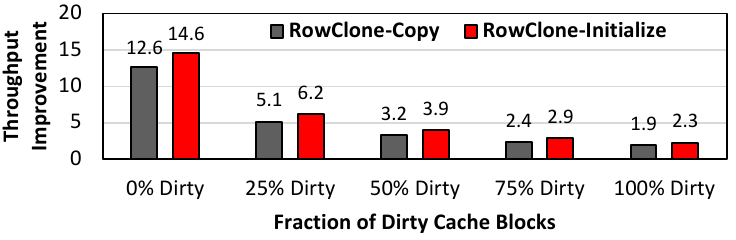}
  \caption{Throughput improvement provided by \texttt{rcc} and \texttt{rci} \revb{with CLFLUSH} over \omi{R}ocket's CPU-copy.}
  \label{fig:system-flush-overhead}
\end{figure}

\Copy{R1/3}{We find that our implementation of CLFLUSH takes 45 \new{Rocket core} \omi{clock} cycles to flush a dirty cache block and 6 \new{Rocket core} cycles to \newnew{invalidate} a clean cache block. {We estimate the throughput improvement of \newnew{\texttt{rcc} and \texttt{rci}} including the CLFLUSH overhead. We assume that all cache blocks of the source \newnew{and destination} operand\newnew{s are} cached, and that a \newnew{fraction} of the \newnew{all cached cache blocks} is dirty (quantified on the x-axis). We do not include the overhead of accessing the data \omi{(e.g., by using \omi{\emph{load}} instructions)} \emph{after} \omi{the data} \newnew{gets copied in DRAM}.} Figure~\ref{fig:system-flush-overhead} shows the {estimated} improvement in \newnew{copy and initialization} throughput \newnew{that} \newnew{\texttt{rcc} and \texttt{rci}} \newnew{provide} for 8 MiB arrays.}

We make three major observations. \newnew{First, even with inefficient cache flush operations, \texttt{rcc} and \texttt{rci} provide 3.2$\times{}$ and 3.9$\times{}$ higher throughput over the CPU-copy and \omi{CPU}-initialization operations, assuming 50\% of the cache blocks of the 8 MiB source operand \omi{are} dirty, respectively.}\revdel{ \newnew{Second,} the copy and initialization throughput provided by \texttt{rcc} and \texttt{rci} are 9.4$\times{}$ and 6.1$\times{}$ smaller when we invalidate clean cache blocks of the source and destination operands of the copy/initialization operations, compared to the \emph{No Flush} configuration, where we assume all data is up-to-date in DRAM. The throughput improvement provided by \texttt{rcc} and \texttt{rci} is 12.6$\times{}$ and 14.6$\times{}$ for 8 MiB arrays over the CPU-copy/initialization baselines.} \newnew{Second,} as the \newnew{fraction} of dirty cache blocks \newnew{increase\omi{s}}, the throughput improvement \newnew{provided by both \texttt{rcc} and \texttt{rci}} \newnew{decreases (down to}\revdel{. The improvement in throughput provided by \texttt{rcc} and \texttt{rci} goes down to} 1.9$\times{}$ for \texttt{rcc} and 2.3$\times{}$ \newnew{for \texttt{rci} \omi{for 100\% dirty cache block \omu{fraction}})}. Third, we observe that \texttt{rci} can provide better throughput improvement compared to \texttt{rcc} when we \omi{include the CLFLUSH overhead. This is because} \texttt{rci} flushes cache blocks of one operand (destination), whereas \texttt{rcc} flushes cache blocks of both operands (source and destination). 

\Copy{R1/4}{{We do not study the distribution of dirty cache block \omi{fractions} in real application\omi{s} as \omi{that} is not the goal of our CLFLUSH overhead analysis. However, if a large dirty {cache block} \omi{fraction} causes severe overhead in a real application, the system designer \newnew{or the user of the system} would likely decide not to offload the operation to PuM (i.e., performing \newnew{\texttt{rcc}} operations instead of CPU-Copy). PiDRAM's prototype can be useful for studies on different PuM system integration aspects, including such offloading decisions.}}

\omi{We observe that the CLFLUSH operations are inefficient in supporting coherence for RowClone operations. Even so, we see that RowClone-Copy and RowClone-Initialization provides throughput improvements ranging from 1.9$\times{}$ to 14.6$\times{}$. We expect the throughput improvement benefits to increase as coherence between the CPU caches and PIM accelerators become more efficient with new techniques~\cite{boroumand2019conda,boroumand2016pim,seshadri2014dirty}.}

\subsubsection{{Real Workload Study}}
\label{sec:real-workload-study}
{The benefit of \texttt{rcc} and \texttt{rci} on a full application depends on what fraction of execution time is spent on bulk data copy and initialization. We demonstrate the benefit of \texttt{rcc} and \texttt{rci} on \emph{forkbench}~\cite{seshadri2013rowclone} and \emph{compile}~\cite{seshadri2013rowclone} workloads with varying fractions of time spent on bulk data copy and initialization, to show that our infrastructure can enable end-to-end execution and estimation of benefits \omi{on real workloads}.\footnote{A full workload study \omi{(i.e., with system calls to a full operating system such as Linux)} of \emph{forkbench} and \emph{compile} is out of the scope of this paper. Our infrastructure currently cannot execute all \omi{possible} workloads due to \omu{the} limited library and system call functionality provided by the \omi{RISC-V Proxy Kernel~\cite{riscv-pk}}.} We especially study \emph{forkbench} in detail to demonstrate how the benefits vary with the time spent on data copying in the baseline for this workload.}

{\emph{Forkbench} first allocates N memory pages and copies data to these pages from a buffer in the process's memory and then accesses 32K random cache blocks within the newly allocated pages to emulate a workload that frequently spawns new processes. We evaluate \emph{forkbench} under varying bulk data copy sizes where we sweep N from 8 to {2048} in increasing powers of two. 

\emph{Compile} first zero-allocates (\texttt{calloc} or \texttt{rci}) two pages (\omi{8 KiBs}) and then executes a number of arithmetic and memory instructions to operate on the zero-allocated data. We carefully develop the \emph{compile} microbenchmark to maintain a realistic ratio between the number of arithmetic and memory instructions executed and zero-allocation function calls made, which we obtain by profiling \emph{gcc}~\cite{perfLinux}. We use the \emph{No-Flush} configuration of our RowClone implementation \omi{for both \emph{forkbench} and \emph{compile}}.}

\begin{figure}[!h]
    \centering
    \includegraphics[width=1.0\linewidth]{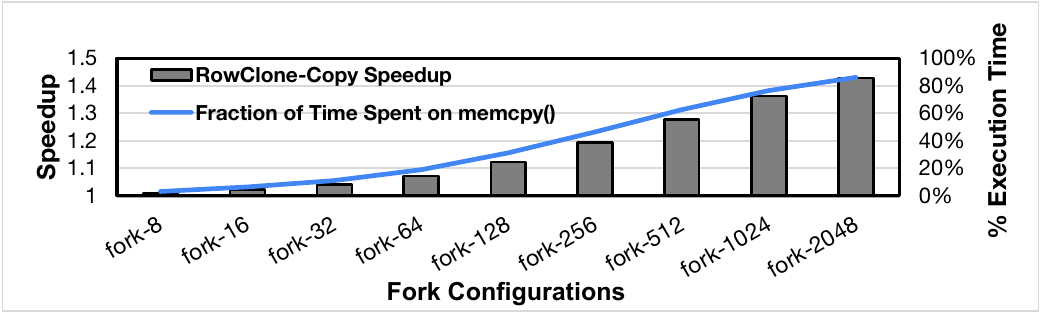}
    \caption{\emph{Forkbench} speedup (bars, left y-axis) and time spent on \texttt{memcpy} by the CPU baseline (curve, right y-axis)}
    \label{fig:fork-all}
\end{figure}

{Figure~\ref{fig:fork-all} plots the speedup provided by \texttt{rcc} over \omi{the} CPU-copy (bars, left y-axis) baseline, and the proportion of time spent on \texttt{memcpy} functions by the CPU-copy baseline (blue curve, right y-axis), for various configurations of \emph{forkbench} on the x-axis.}

\noindent
{\textbf{Forkbench.}} {We observe that RowClone-Copy can significantly improve the performance of \emph{forkbench} by up to {42.9}\%. \omi{RowClone-Copy's} performance improvement increases as the number of pages copied increase. This is because the copy operations accelerated by \texttt{rcc} contribute a larger amount to the total execution time of the workload. The \texttt{memcpy} function calls take {86\%} of the CPU-copy baseline's time during \emph{forkbench} execution for N = {2048}.}

\noindent
{\textbf{Compile.}} {RowClone-Initialize \omi{improves} the performance of \emph{compile} by 9\%. Only an estimated 17\% of the execution time of \emph{compile} is used for zero-allocation by the CPU-initialization baseline, \texttt{rci} reduces the overhead of zero-allocation by \omi{(i)} performing \omi{in-DRAM} bulk-initialization and \omi{(ii)} executing a smaller number of instructions.}

\noindent
{\textbf{Libquantum.} To demonstrate that \X can run real workloads, we run a SPEC2006~\cite{spec2006} workload (libquantum). We modify the \texttt{calloc} (allocates and zero initializes memory) function call to allocate data using \texttt{alloc\_align}, and initialize data using \texttt{rci} for allocations that are larger than 8 KiBs.}

\revcommon{Using \texttt{rci} to bulk initialize data in libquantum improves end-to-end application performance by 1.3\% (compared to the baseline that uses CPU-Initialization). This improvement is brought by \texttt{rci}, which initializes a total amount of 512 KiBs of memory\footnote{\revcommon{\omi{In libquantum,} there are 16 calls to \texttt{calloc} that exceed the 8 KiB allocation size. \omi{W}e only bulk initialize data using \texttt{rci} for these 16 calls.}} using RowClone operations.} {We note that the proportion of store instructions executed by libquantum to initialize arrays in the CPU-initialization baseline is only 0.2\% of all dynamic instructions in the libquantum workload {which amounts to an estimated 2.3\% of the total runtime of libquantum.} \omi{T}hus, the \omi{1.3\% end-to-end} performance improvement provided by \texttt{rci}, which can \omi{ideally} speed up only 2.3\% of \omu{the} total runtime, is reasonable, and we expect it to increase with the initialization intensity of workloads.}

\noindent
\omi{\textbf{Summary.}}
\revd{We conclude from our {evaluation} that end-to-end implementations of RowClone (i) can be \omi{efficiently} supported in real systems by employing memory allocation mechanisms that satisfy the memory \emph{alignment}, \emph{mapping}, \emph{granularity} requirements (Section~\ref{sec:rowclone_alignment}) of RowClone operations, (ii) can greatly improve copy/initialization throughput in real systems, and (iii) require cache coherenc\omi{e} mechanisms (e.g., PIM-optimized coherenc\omi{e} management~\cite{boroumand2019conda,boroumand2016pim,seshadri2014dirty}) that can flush dirty cache blocks of RowClone operands efficiently to achieve optimal copy/initialization throughput improvement.} {\X{} can be used \omi{to} estimat\omi{e} end-to-end \omi{workload} execution benefits provided by RowClone operations.} \revcommon{Our experiment\omi{s} using libquantum\omi{, forkbench, and compile} show that (i) \X can run real workloads, (ii) our end-to-end implementation of RowClone operates correctly, and (iii) RowClone can improve the performance of real workloads \omi{in a real system, even when inefficient CLFLUSH operations are used to maintain memory coherence}.}

\section{Case Study \#2: End-to-end D-RaNGe}

\label{sec:drange}

{Prior work on DRAM-based random number generation techniques~\cite{olgun2021quactrng,kim.hpca19,talukder2019exploiting} do not integrate and evaluate their techniques end-to-end in a real system.}
\revdel{D-RaNGe~\cite{kim.hpca19} (Section~\ref{sec:background_pudram}) {is a state-of-the-art} DRAM-based true random number generat{ion technique} that leverages the randomness in DRAM activation latency ($tRCD$) failures.} \new{We evaluate one DRAM-based true random number generation technique, D-RaNGe~\cite{kim.hpca19}, end-to-end using PiDRAM.} We implement support for D-RaNGe in PiDRAM by enabling \omi{access to DRAM with} \new{reduced activation latency} (i.e., \omi{$tRCD$ set to values lower than} manufacturer recommendations).

\subsection{D-RaNGe Implementation}

We implement a simple version of D-RaNGe in PiDRAM. PiDRAM's D-RaNGe \omi{controller} collects true random numbers from four DRAM cells in the same DRAM cache block 
\omi{inside} one DRAM bank. 
We implement the \omi{D-RaNGe controller} within the \revf{Periodic Operations Module \new{(Section~\ref{sec:hardware-components})}}. The \omi{D-RaNGe controller} (i) periodically accesses a DRAM cache block with reduced tRCD, (ii) reads four of the TRNG DRAM cells in the cache block, (iii) stores the four bits read from the TRNG cells in a 1 KiB \new{random number buffer}. We reserve multiple configuration registers in the configuration register file (CRF) to configure (i) the \omi{TRNG period (in nanoseconds) used by} the \omi{D-RaNGe controller} to periodically generate random numbers \omi{by accessing DRAM with reduced activation latency} while the buffer is not full (the D-RaNGe controller accesses DRAM every TRNG period), (ii) the timing parameter ($tRCD$) used \omi{to} induc\omi{e} activation latency failures and (iii) the \new{physical} location \new{(DRAM bank, row, column addresses, and bit offset within the DRAM column)} of the TRNG cells \omi{to read}. We implement two pumolib functions: (i) \texttt{buf\_\omi{size}()}, which returns the number of random words (4-bytes) available in the buffer, and (ii) \texttt{rand\_dram()}, which returns one random word that is read from the buffer. {The two functions \new{first} execute PiDRAM instructions in the POC that \new{update} the data register either with (i) the number of random words available \new{(when buf\_s\omi{ize}() is called)} or (ii) a random word read from the random number buffer \new{(when rand\_dram() is called)}. The two functions \new{then} access the data register using LOAD instructions to retrieve \omi{either} the size of the random number buffer or a random number.} The application developer reads true random numbers using these two functions in pumolib.

\noindent
\textbf{\new{Random Cell Characterization.}} D-RaNGe requires the system designer to characterize the DRAM module for activation latency failures to find DRAM cells that fail with a 50\% probability \omi{(i.e., randomly)} when accessed with reduced $tRCD$. \revd{Following the methodology presented in~\cite{kim.hpca19}, the system designer can characterize a DRAM device or use an automated procedure to find cells that fail with a 50\% probability.} \omi{In PiDRAM,} we implement reduced latency access to DRAM by (i) extending the scheduler of the custom memory controller and (ii) adding a pumolib function \texttt{\omi{activation\_failure}(\omi{address})} which induces an activation failure on the DRAM cache block pointed by the \texttt{\omi{address}} parameter.

\subsection{Evaluation and Results}
\label{sec:drange-evaluation}
\textbf{Experimental Methodology.} We run a microbenchmark to understand the effect of the TRNG period on true random number \omi{generation} throughput observed by a program running on the \omi{R}ocket core. The microbenchmark consists of a loop that (i) checks the availability of random numbers using \texttt{buf\_\omi{size}()} and (ii) reads a random number from the buffer using \texttt{rand\_dram()}. \newnew{We execute the microbenchmark until we read one million bytes of random numbers.} 

\begin{figure}[!ht]
  \centering
  \vspace{-5mm}
  \includegraphics[width=0.40\textwidth]{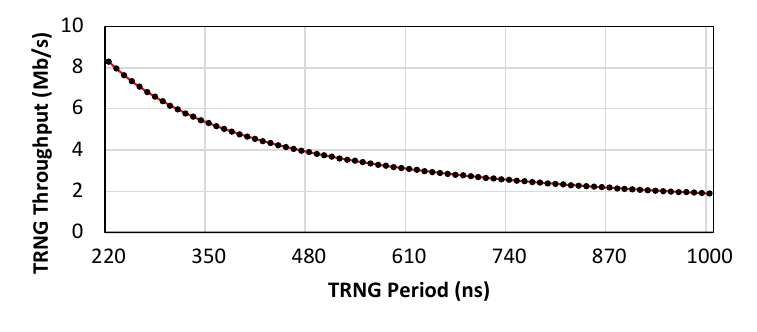}
  \vspace{-3mm}
  \caption{TRNG throughput observed by our microbenchmark for TRNG periods ranging from 220 $ns$ to 1000 $ns$}
  \label{fig:trng-throughput}
\end{figure}

\noindent
\textbf{Results.} \omi{The D-RaNGe controller} can perform reduced-latency accesses frequently, every 220 $ns$. Figure~\ref{fig:trng-throughput} depicts the TRNG throughput observed by the microbenchmark for TRNG periods in the range [220 $ns$, 1000 $ns$] with increments of 10 $ns$. We observe that the TRNG throughput decreases from 8.30 Mb/s at 220 $ns$ TRNG period to 1.90 Mb/s at 1000 $ns$ TRNG period. D-RaNGe~\cite{kim.hpca19} reports 25.2 Mb/s TRNG throughput using a single DRAM bank when there are \new{four} random cells in a cache block. PiDRAM's \omi{D-RaNGe controller} can be optimized to generate random numbers more frequently to match D-RaNGe's observed maximum throughput.\footnote{\omi{D-RaNGe has a smaller true random number generation (TRNG) latency (i.e., takes a smaller amount of time to generate a 4-bit random number) than PiDRAM. PiDRAM has a larger TRNG latency due to (i) discrepancies in the data path (i.e., on-chip interconnect) in D-RaNGe's simulated system and PiDRAM's prototype and (ii) \omu{the TRNG period of the D-RaNGe controller (D-RaNGe controller performs a reduced $tRCD$ access only as frequently as one every \SI{220}{\nano\second}).} \omu{The D-RaNGe controller} can be optimized further to reduce the TRNG period \omu{by} down to the DRAM row cycle time ($tRC$ standard timing parameter, typically ~\SI{45}{\nano\second}~\cite{micron2018ddr3}).}} We leave such optimizations to PiDRAM's \omi{D-RaNGe controller} for future work.


Including the modifications to the custom memory controller and pumolib, implementing D-RaNGe and reduced-latency DRAM access \new{requires} an additional \textbf{190} lines of Verilog and \textbf{74} lines of C code over \X's existing codebase. We conclude that our D-RaNGe implementation (i) provides a basis for \X developers to study end-to-end implementations of \omi{DRAM-based true random number generators}, (ii) shows that \X's hardware and software components facilitate the implementation of new {commodity DRAM based} PuM techniques\omi{, specifically those that are related to security}. Our reduced-latency DRAM access implementation provides a basis for other PuM techniques \omi{for security purposes}, such as \omi{the DRAM-latency physical unclonable functions (}DL-PUF~\cite{kim.hpca18}) \omi{and QUAC-TRNG~\cite{olgun2021quactrngieee} (Section~\ref{sec:use-cases})}. We leave further exploration on end-to-end implementations of D-RaNGe\omi{, DL-PUF, and QUAC-TRNG, as well as end-to-end analyses of \omu{the} security benefits they provide using PiDRAM} for future work.

\section{{Extending PiDRAM}}
\label{sec:extending-pidram}
{We briefly describe the modifications required to extend PiDRAM (i) with new DRAM commands and DRAM timing parameters\new{, (ii) with new case studies,} and (i\new{i}i) to support new FPGA boards.}

\noindent
\textbf{{New DRAM Commands and Timing Parameters.}}
\Copy{R1/1}{{Implementing new DRAM commands or modifying DRAM timing parameters require modifications to PiDRAM's memory controller. This is straightforward as PiDRAM's memory controller's Verilog design \omi{is modular and uses well-defined interfaces: It is composed} of multiple modules that perform separate tasks\revdel{and communicate with each other via well-defined interfaces}. For example, the memory request scheduler comprises \new{two} main components: (1) \emph{command timer}, \new{and} (2) \emph{command scheduler}\new{.}\revdel{, and (3) custom command scheduler} To serve {LOAD and STORE memory} requests, the command scheduler maintains state (e.g., which row is active) for every bank. The command scheduler selects the next DRAM command to satisfy the {LOAD or STORE memory} request and queries the command timer with the selected DRAM command. The command timer checks for all possible standard DRAM timing constraints and outputs a valid bit if the selected command can be issued in that FPGA clock cycle. To extend the memory controller with a new standard DRAM command (e.g., to implement a newer standard like DDR4 \omi{or DDR5}), a PiDRAM developer simply needs to \omi{(i)} add a new timing constraint by replicating the logic in the command timer and \omi{(ii)} extend the command scheduler to correctly maintain the bank state.}}

\noindent
{\textbf{New Case Studies.}}
\Copy{R5/2}{{Implementing new techniques (e.g., those that are listed in Table~\ref{table:use-cases}) to perform new case studies requires modifications to PiDRAM's hardware and software components. We describe the required modifications over an example {ComputeDRAM-based in-DRAM bitwise operations} case study.}

{To implement ComputeDRAM-based in-DRAM bitwise operations, the developers need to (i) extend the \emph{\new{custom command} scheduler} in PiDRAM's memory controller with a new state machine that schedules new DRAM command sequences (ACT-PRE-ACT) with an appropriate set of violated timing parameters (our ComputeDRAM-based in-DRAM copy implementation provides a \omi{solid} basis for this), (ii) expose the functionality to the processor by implementing new PiDRAM instructions in the PuM controller (e.g., by replicating \omi{and customizing} the existing logic for decoding and executing RowClone operations), (iii) and make modifications to the software library to expose the new instruction to the programmer (e.g., by replicating the copy\_row function's behavior, described in Table~\ref{table:pumolib}).}}

\noindent
\textbf{{Porting to New FPGA Boards.}}
\Copy{R3/7A}{{Developing new PiDRAM prototypes on different FPGA boards could require modifications to design constraints (e.g., top level input/outputs to physical FPGA pins) and the DDRx PHY IP depending on the FPGA board. Modifying design constraints is a straightforward task involving looking up the FPGA manufacturer datasheets and modifying design constraint files~\cite{designconstraints}. Manufacturers may provide different DDRx PHY IPs for different FPGAs. Fortunately, these IPs typically expose \omi{similar} (based on the DFI standard~\cite{dfi}) interface\omi{s} to user hardware (in our case, to PiDRAM's memory controller). Thus, \onur{other} PiDRAM prototypes on different FPGA boards can be developed with \omi{small yet careful} modifications to the ZC706 prototype design we provide.}}

\section{Related Work}
\label{sec:related-work}

To our knowledge, this is the first work to develop a flexible, open-source framework that enables integration and evaluation of \new{commodity DRAM based} \omi{processing-using-memory} (PuM) techniques on real DRAM chips by providing the necessary hardware and software components. 
We demonstrate the first end-to-end implementation of RowClone \omi{and D-RaNGe using real DRAM chips}. We compare the features of \X with other state-of-the-art \omi{prototyping and evaluation} platforms in Table~\ref{table:tools} and discuss them below. \omi{The four features we use for comparison are:} 

\begin{enumerate}
    \item \textbf{Interface with real DRAM chips:} The platform allows running experiments using real DRAM chips.
    \item \textbf{Flexible memory controller (MC) for PuM:} The platform provides a flexible memory controller that can easily be extended to perform (e.g., as in PiDRAM) or emulate (e.g., as in PiMulator~\cite{mosanu2022pimulator}) new PuM operations.
    \item \textbf{System software support:} The platform provides support for running system software such as operating systems or supervisor software (e.g., RISC-V PK~\cite{riscv-pk}).
    \item \textbf{Open-source:} The platform is available as open source software.
\end{enumerate}

\begin{table*}[!t]
\centering
\caption{Compari\omi{son of} \X with related state-of-the-art \omi{prototyping and evaluation} platforms}
\label{table:tools}
\resizebox{1.0\textwidth}{!}{\begin{tabular}{|l||c|c|c|c|}

\hline
           \textbf{Platforms} & \begin{tabular}[c]{@{}l@{}}\textbf{Interface with \omi{real} DRAM \omi{chips}}\end{tabular} & \begin{tabular}[c]{@{}l@{}}\textbf{Flexible MC for PuM}\end{tabular} & \begin{tabular}[c]{@{}l@{}}
           \textbf{System software support}\end{tabular}  & \begin{tabular}[c]{@{}l@{}}\textbf{Open-source}\end{tabular}  \\ \hline \hline
\textbf{Silent-PIM} \cite{kim2021SilentPIM} & \xmark & \xmark & \checkmark  & \xmark\\ \hline
\textbf{SoftMC}  \cite{hassan2017softmc}   & \checkmark (DDR3) &  \xmark  &  \xmark &  \checkmark\\ \hline
\textbf{ComputeDRAM}   \cite{gao2020computedram}    & \checkmark (DDR3) & \xmark & \xmark   & \xmark \\ \hline
\textbf{MEG}    \cite{zhang2020MEG}    & \checkmark (HBM) &  \xmark  & \checkmark  &   \checkmark  \\ \hline
{\textbf{PiMulator}}    \cite{mosanu2022pimulator}    & \xmark &  \checkmark  & \xmark  &   \checkmark  \\ \hline
\textbf{\omi{Commercial platforms (e.g.,} ZYNQ~\cite{zynq})} & \checkmark (DDR3/4) &  \xmark    & \checkmark   &  \xmark \\ \hline
\textbf{Simulators}  \cite{gem5-gpu,GEM5,ramulator,ramulator-pim,zhang2022pim,forlin2022sim2pim,yu2021multipim,xu2019pimsim} & \xmark &  \checkmark  &  \checkmark \omi{(potentially)}   & \checkmark  \\ \hline
\hline
\textbf{\X  (this work)}   & \checkmark (DDR3) &  \checkmark   &  \checkmark &  \checkmark  \\ \hline
\end{tabular}}

\end{table*}

\atb{\textbf{Silent-PIM~\cite{kim2021SilentPIM}.} Silent-PIM proposes a new DRAM \new{design} \revdel{(similar to FIMDRAM~\cite{kwon2021fimdram}) }that incorporates processing units capable of vector arithmetic computation. \revdel{The authors use the standard DRAM interface~\cite{jedecDDR4} to communicate with the DRAM device. }Silent-PIM's goal is to evaluate PIM techniques on a \emph{new, PIM-capable} DRAM device using standard DRAM commands \omi{(e.g., as defined in DDR4~\cite{jedecDDR4})}\omi{; it does not \omu{provide} an evaluation platform or prototype}. \atb{\new{In contrast}, \X is designed for researchers to rapidly integrate and evaluate PuM techniques that use \emph{real DRAM devices}. \X provides key hardware and software components that facilitate end-to-end implementations of PuM techniques.}}

\revb{\textbf{SoftMC~\cite{hassan2017softmc, softmc.github}} SoftMC is an FPGA-based DRAM testing infrastructure. SoftMC \revb{can issue} arbitrary sequences of DDR3 commands to real DRAM modules. {SoftMC is widely used in prior work that \omi{studies} the performance, reliability and security of real DRAM chips~\cite{frigo2020trr, hassan2021uncovering, kim2020revisiting, khan.micro17, chang.hpca16, talukder2020towards, farmani2021RHAT, kim.hpca19, talukder2019exploiting, lee.sigmetrics17, ghose2018vampire, orosaYaglikci2021deeper,talukder2019prelatpuf}}. \Copy{R5/2B}{SoftMC is built to test DRAM modules, \emph{not} to study end-to-end implementations of PuM techniques. \new{Thus,} SoftMC (i) does \emph{not} \omi{support application execution on a real system}, and (ii) \revb{\emph{cannot} use} DRAM modules as \revb{{main memory}}. {While SoftMC is useful in studies that perform exhaustive search on all possible sequences of DRAM commands to potentially uncover undocumented DRAM behavior (e.g., ComputeDRAM~\cite{gao2020computedram}, QUAC-TRNG~\cite{olgun2021quactrng}),} \X is developed to study end-to-end implementations of PuM techniques. \X provides an FPGA-based prototype that comprises a RISC-V system and supports using DRAM modules both for storing data \omi{(i.e., as main memory)} and \omi{performing PuM} computation.}}

\textbf{\atb{ComputeDRAM~\cite{gao2020computedram}.}} ComputeDRAM partially demonstrates \omi{that} two DRAM-based state-of-the-art PuM techniques\new{,} RowClone~\cite{seshadri2013rowclone} and Ambit~\cite{seshadri.micro17}\new{, \omi{are already possible} on real \omi{off-the-shelf} DDR3 chips}. ComputeDRAM uses SoftMC to demonstrate in-DRAM copy and bitwise \omi{AND/OR} operations on real DDR3 chips. ComputeDRAM's goal is \emph{not} to develop a framework to facilitate end-to-end implementations of PuM techniques. Therefore, it does \emph{not} provide (i) a flexible memory controller for PuM \omi{or}, (ii) support for system software. \X provides the necessary software and hardware components to facilitate end-to-end implementations of PuM techniques.

\textbf{MEG~\cite{zhang2020MEG}.} MEG is an open-source system emulation platform for enabling FPGA-based operation interfacing with High-Bandwidth Memory (HBM). MEG aims to efficiently retrieve data from HBM and perform the computation in the host processor implemented as a soft core on the FPGA. Unlike \X, MEG does \emph{not} implement a flexible memory controller that is capable of performing PuM operations. We demonstrate the flexibility of \X by implementing two state-of-the-art PuM techniques~\cite{seshadri2013rowclone,kim.hpca19}. \omi{We believe MEG and PiDRAM can be combined to get the functionality and prototyping power of both works.}

{\textbf{PiMulator~\cite{mosanu2022pimulator}.} PiMulator is an open-source PiM emulation platform. PiMulator implements a main memory and a PiM model using SystemVerilog, allowing FPGA emulation of PiM architectures. PiMulator enables easy emulation of new PiM techniques. However, it does \emph{not} allow end-to-end execution of workloads that use PiM techniques and it does not provide the user \omi{with} full control over the DRAM interface.}

\textbf{Commercial Platforms (e.g., ZYNQ \cite{zynq}).} Some commercial platforms implement CPU-FPGA heterogeneous computing systems. A memory controller and necessary hardware-software modules are provided to access DRAM as the main memory in such systems. However, in such systems, (i) there is \emph{no} support for PuM mechanisms\new{, and} (ii) the entire hardware-software stack is closed-source. \X can be integrated into these systems, using the closed-source computing system as the main processor. Our prototype utilizes an open-source system-on-chip (\omi{Rocket Chip}~\cite{asanovic2016rocket}) as the main processor, which enables developers to study architectural \omi{and microarchitectur\omu{al}} aspects of PuM techniques (e.g., \omi{data allocation and} coherence mechanisms). Such studies cannot be conducted \omi{using} closed-source computing systems.

\textbf{Simulators.} Many prior works propose full-system \omu{(e.g.,~\cite{GEM5,gem5-gpu})}, trace-based \omu{(e.g.,~\cite{ramulator-pim,ramulator,zhang2022pim,yu2021multipim,xu2019pimsim,scarab})}, and instrumentation-based \omu{(e.g.,~\cite{scarab,forlin2022sim2pim,xu2019pimsim})} simulators that can be used to evaluate PuM techniques. Although useful, these simulators do not model DRAM behavior and cannot integrate proprietary device characteristics (e.g., DRAM internal address mapping) into their simulations, \atb{without conducting a rigorous characterization study. Moreover, the effects of environmental conditions (e.g., temperature, voltage) on DRAM chips are unlikely to be modeled on \revcommon{accurate, full-system} simulators as it would require excessive computation, \revcommon{which} would \revcommon{negatively impact the} \revcommon{already poor} performance \revcommon{(200K instructions per second)} of \revcommon{full system simulators} ~\cite{zsim}}. \atb{\omi{In contrast}, \X interfaces with real DRAM devices \revcommon{and its prototype \omi{achieves} a 50 MHz clock speed (\omi{and} can be improved \omi{further}) which lets \X execute > 10M instructions per second (assuming < 5 cycles per instruction)}. \Copy{R3/5}{\X can be used to study end-to-end implementations of PuM techniques and explore solutions that take into account \revcommon{the effects related to the environmental conditions of real DRAM devices. {Future versions of \X{} could be easily extended (e.g., with real hardware that allows controlling DRAM temperature and voltage~\cite{pl068p,maxwellFT200}) to experiment with different DRAM temperature and voltage levels to better understand the effects of these environmental conditions on the reliability of PuM operations.} Using \X, experiments that require executing real workloads can take an order of magnitude \omi{shorter} wall clock time compared to using full-system simulators.}}}

{\textbf{Other Related Work.} \omi{Prior works (see Section~\ref{sec:background_pudram}) (i) propose or (ii) demonstrate using real DRAM chips, several DRAM-based PuM techniques that can perform computation~\cite{Seshadri:2015:ANDOR, seshadri.arxiv16, seshadri.micro17, seshadri.bookchapter17, seshadri2020indram,hajinazarsimdram,chang.hpca16,ferreira2021pluto,angizi2019graphide}, move data~\cite{seshadri2013rowclone,wang2020figaro}, or implement security primitives~\cite{olgun2021quactrngieee,kim.hpca18,kim.hpca19,talukder2019exploiting,talukder2019prelatpuf,orosa2021codic} in memory.}\revdel{{Seshadri et al.~\cite{Seshadri:2015:ANDOR, seshadri.arxiv16, seshadri.micro17, seshadri.bookchapter17, seshadri2020indram}} propose using triple row activation in DRAM to perform bitwise majority \omi{(and thus AND/OR) and NOT} operations across the three activated rows. ComputeDRAM~\cite{gao2020computedram} shows that a subset of real DDR3 chips can perform triple row activation when they receive an ACT-PRE-ACT command sequence with violated timing parameters.} SIMDRAM~\cite{hajinazarsimdram} develops a framework that provides a programming interface to perform in-DRAM computation using the majority operation. \omi{DR-STRANGE~\cite{bostanci2022drstrange} proposes an end-to-end system design for DRAM-based true random number generators.} None of these works provide an end-to-end in-DRAM computation framework that is integrated into a real system \omu{using real DRAM chips}.}

{We conclude that} existing platforms cannot substitute PiDRAM in studying \new{commodity DRAM based} PuM techniques end-to-end.


\section{Conclusion}
We develop \X, a flexible and open-source \omi{prototyping} framework for integrating and evaluating end-to-end {commodity DRAM based} \omi{processing-using-memory} (PuM) techniques.
\X comprises the necessary \omi{hardware and software} structures to facilitate end-to-end implementation of PuM techniques.
We build an FPGA-based prototype of \X along with an open-source RISC-V system and enable computation on real DRAM chips.
Using \X, we implement and evaluate RowClone \omi{(in-DRAM data copy and initialization)} and D-RaNGe \omi{(in-DRAM true random number generation)} end-to-end \omi{in the entire real system}. 
Our results show that RowClone significantly improve\omi{s} data copy \omi{and initialization} throughput \omi{in a real system on real workloads}, and efficient cache coherence mechanisms are \omu{needed} 
to maximize \omi{RowClone's potential benefits}. Our implementation of D-RaNGe requir\omi{es small} additions to \omi{\X{}'s codebase and} \omi{provides true random numbers at high throughput and with low latency}.
We conclude that \omi{unlike existing prototyping and evaluation platforms,} \X enables (i) easy integration of \omi{existing and new} PuM techniques \omi{end-to-end in a real system} and (ii) novel studies on end-to-end implementations of PuM techniques \omi{using real DRAM chips}. \new{PiDRAM is \omi{freely} available} as an open-source tool \omu{for} researchers and \omi{designers} \omi{in both academia and industry to experiment with and build on.}

\section*{Acknowledgements}
This research was partially supported by ACCESS – AI Chip Center for Emerging Smart Systems, sponsored by InnoHK funding, Hong Kong SAR.

\balance
\begin{spacing}{0.75}
\begin{footnotesize}
\bibliographystyle{IEEEtranS}
\bibliography{PiDRAM}

\begin{thebibliography}{100}
\providecommand{\url}[1]{#1}
\csname url@samestyle\endcsname
\providecommand{\newblock}{\relax}
\providecommand{\bibinfo}[2]{#2}
\providecommand{\BIBentrySTDinterwordspacing}{\spaceskip=0pt\relax}
\providecommand{\BIBentryALTinterwordstretchfactor}{4}
\providecommand{\BIBentryALTinterwordspacing}{\spaceskip=\fontdimen2\font plus
\BIBentryALTinterwordstretchfactor\fontdimen3\font minus
  \fontdimen4\font\relax}
\providecommand{\BIBforeignlanguage}[2]{{%
\expandafter\ifx\csname l@#1\endcsname\relax
\typeout{** WARNING: IEEEtranS.bst: No hyphenation pattern has been}%
\typeout{** loaded for the language `#1'. Using the pattern for}%
\typeout{** the default language instead.}%
\else
\language=\csname l@#1\endcsname
\fi
#2}}
\providecommand{\BIBdecl}{\relax}
\BIBdecl

\bibitem{aga.hpca17}
S.~Aga, S.~Jeloka, A.~Subramaniyan, S.~Narayanasamy, D.~Blaauw, and R.~Das,
  ``{Compute Caches},'' in \emph{HPCA}, 2017.

\bibitem{ahn.tesseract.isca15}
J.~Ahn, S.~Hong, S.~Yoo, O.~Mutlu, and K.~Choi, ``{A Scalable
  Processing-in-Memory Accelerator for Parallel Graph Processing},'' in
  \emph{ISCA}, 2015.

\bibitem{ahn.pei.isca15}
J.~Ahn, S.~Yoo, O.~Mutlu, and K.~Choi, ``{PIM-Enabled Instructions: A
  Low-Overhead, Locality-Aware Processing-in-Memory Architecture},'' in
  \emph{ISCA}, 2015.

\bibitem{DBLP:conf/isca/AkinFH15}
B.~Akin, F.~Franchetti, and J.~C. Hoe, ``{Data Reorganization in Memory Using
  {3D}-Stacked {DRAM}},'' in \emph{ISCA}, 2015.

\bibitem{ali2019memory}
M.~F. Ali, A.~Jaiswal, and K.~Roy, ``{In-Memory Low-Cost Bit-Serial Addition
  Using Commodity DRAM Technology},'' in \emph{{TCAS-I}}, 2019.

\bibitem{angizi2018pima}
S.~Angizi, Z.~He, and D.~Fan, ``{PIMA-Logic: A Novel Processing-in-Memory
  Architecture for Highly Flexible and Energy-efficient Logic Computation},''
  in \emph{DAC}, 2018.

\bibitem{angizi2018cmp}
S.~Angizi, A.~S. Rakin, and D.~Fan, ``{CMP-PIM: An Energy-efficient
  Comparator-based Processing-in-Memory Neural Network Accelerator},'' in
  \emph{DAC}, 2018.

\bibitem{angizi2019dna}
S.~Angizi, J.~Sun, W.~Zhang, and D.~Fan, ``{AlignS: A Processing-in-Memory
  Accelerator for DNA Short Read Alignment Leveraging SOT-MRAM},'' in
  \emph{DAC}, 2019.

\bibitem{angizi2019graphide}
S.~Angizi and D.~Fan, ``{Graphide: A Graph Processing Accelerator Leveraging
  In-DRAM-Computing},'' in \emph{GLSVLSI}, 2019.

\bibitem{arm-cmos}
\BIBentryALTinterwordspacing
ARM, ``{Cache Maintenance Operations},'' 2021. [Online]. Available:
  \url{https://developer.arm.com/documentation/ddi0246/h/programmers-model/register-descriptions/cache-maintenance-operations}
\BIBentrySTDinterwordspacing

\bibitem{asanovic2016rocket}
K.~Asanovi{\'c}, R.~Avizienis, J.~Bachrach, S.~Beamer, D.~Biancolin, C.~Celio,
  H.~Cook, P.~Dabbelt, J.~R. Hauser, A.~M. Izraelevitz, S.~Karandikar,
  B.~Keller, D.~Kim, J.~Koenig, Y.~Lee, E.~Love, M.~Maas, A.~Magyar, H.~Mao,
  M.~Moret{\'o}, A.~Ou, D.~A. Patterson, B.~H. Richards, C.~Schmidt, S.~M.
  Twigg, H.~Vo, and A.~Waterman, ``The rocket chip generator,'' ser. Technical
  Report No. UCB/EECS-2016-17, 2016.

\bibitem{asghari-moghaddam.micro16}
H.~Asghari-Moghaddam, Y.~H. Son, J.~H. Ahn, and N.~S. Kim, ``{Chameleon:
  Versatile and Practical Near-DRAM Acceleration Architecture for Large Memory
  Systems},'' in \emph{MICRO}, 2016.

\bibitem{talukder2019exploiting}
B.~M.~S. {Bahar Talukder}, J.~{Kerns}, B.~{Ray}, T.~{Morris}, and M.~T.
  {Rahman}, ``{Exploiting DRAM Latency Variations for Generating True Random
  Numbers},'' in \emph{ICCE}, 2019.

\bibitem{talukder2019prelatpuf}
B.~M.~S. Bahar~Talukder, B.~Ray, D.~Forte, and M.~T. Rahman, ``{PreLatPUF:
  Exploiting DRAM Latency Variations for Generating Robust Device
  Signatures},'' in \emph{IEEE Access}, 2019.

\bibitem{barenghi2018software}
A.~Barenghi, L.~Breveglieri, N.~Izzo, and G.~Pelosi, ``{Software-only Reverse
  Engineering of Physical DRAM Mappings for Rowhammer Attacks},'' in
  \emph{IVSW}, 2018.

\bibitem{besta2021sisa}
M.~Besta, R.~Kanakagiri, G.~Kwasniewski, R.~Ausavarungnirun, J.~Ber\'{a}nek,
  K.~Kanellopoulos, K.~Janda, Z.~Vonarburg-Shmaria, L.~Gianinazzi, I.~Stefan,
  J.~G. Luna, J.~Golinowski, M.~Copik, L.~Kapp-Schwoerer, S.~Di~Girolamo,
  N.~Blach, M.~Konieczny, O.~Mutlu, and T.~Hoefler, ``{SISA: Set-Centric
  Instruction Set Architecture for Graph Mining on Processing-in-Memory
  Systems},'' in \emph{MICRO}, 2021.

\bibitem{bhattacharjee2017revamp}
D.~Bhattacharjee, R.~Devadoss, and A.~Chattopadhyay, ``{ReVAMP: ReRAM based
  VLIW Architecture for In-Memory Computing},'' in \emph{DATE}, 2017.

\bibitem{GEM5}
N.~Binkert, B.~Beckman, A.~Saidi, G.~Black, and A.~Basu, ``{The gem5
  Simulator},'' \emph{CAN}, 2011.

\bibitem{borghetti2010memristive}
J.~Borghetti, G.~Snider, P.~Kuekes, J.~J. Yang, D.~Stewart, and S.~Williams,
  ``{Memristive Switches Enable Stateful Logic Operations via Material
  Implication},'' in \emph{Nature}, 2010.

\bibitem{boroumand.asplos18}
A.~Boroumand, S.~Ghose, Y.~Kim, R.~Ausavarungnirun, E.~Shiu, R.~Thakur, D.~Kim,
  A.~Kuusela, A.~Knies, P.~Ranganathan, and O.~Mutlu, ``{Google Workloads for
  Consumer Devices: Mitigating Data Movement Bottlenecks},'' in \emph{ASPLOS},
  2018.

\bibitem{boroumand2019conda}
A.~Boroumand, S.~Ghose, M.~Patel, H.~Hassan, B.~Lucia, R.~Ausavarungnirun,
  K.~Hsieh, N.~Hajinazar, K.~T. Malladi, H.~Zheng, and O.~Mutlu, ``{CoNDA:
  Efficient Cache Coherence Support for near-Data Accelerators},'' in
  \emph{ISCA}, 2019.

\bibitem{boroumand2016pim}
A.~Boroumand, S.~Ghose, M.~Patel, H.~Hassan, B.~Lucia, K.~Hsieh, K.~T. Malladi,
  H.~Zheng, and O.~Mutlu, ``{LazyPIM: An Efficient Cache Coherence Mechanism
  for Processing-in-Memory},'' in \emph{CAL}, 2016.

\bibitem{bostanci2022drstrange}
F.~Bostanci, A.~Olgun, L.~Orosa, A.~Yaglikci, J.~S. Kim, H.~Hassan, O.~Ergin,
  and O.~Mutlu, ``{DR-STRaNGe: End-to-End System Design for DRAM-based True
  Random Number Generators},'' in \emph{HPCA}, 2022.

\bibitem{geoffrey2017neuromorphic}
G.~W. Burr, R.~M. Shelby, A.~Sebastian, S.~Kim, S.~Kim, S.~Sidler, K.~Virwani,
  M.~Ishii, P.~Narayanan, A.~Fumarola, L.~L. Sanches, I.~Boybat, M.~L. Gallo,
  K.~Moon, J.~Woo, H.~Hwang, and Y.~Leblebici, ``{Neuromorphic Computing Using
  Non-volatile Memory},'' in \emph{Advances in Physics: X}, 2017.

\bibitem{cali2020genasm}
D.~S. Cali, G.~S. Kalsi, Z.~Bingöl, C.~Firtina, L.~Subramanian, J.~S. Kim,
  R.~Ausavarungnirun, M.~Alser, J.~Gomez-Luna, A.~Boroumand, A.~Nori,
  A.~Scibisz, S.~Subramoney, C.~Alkan, S.~Ghose, and O.~Mutlu, ``{GenASM: A
  High-Performance, Low-Power Approximate String Matching Acceleration
  Framework for Genome Sequence Analysis},'' in \emph{MICRO}, 2020.

\bibitem{kevinchang-thesis}
K.~Chang, ``{Understanding and Improving the Latency of DRAM-Based Memory
  Systems},'' Ph.D. dissertation, Carnegie Mellon University, 2017.

\bibitem{chang.sigmetrics16}
K.~K. Chang, A.~Kashyap, H.~Hassan, S.~Ghose, K.~Hsieh, D.~Lee, T.~Li,
  G.~Pekhimenko, S.~Khan, and O.~Mutlu, ``{Understanding Latency Variation in
  Modern DRAM Chips: Experimental Characterization, Analysis, and
  Optimization},'' in \emph{SIGMETRICS}, 2016.

\bibitem{chang.hpca16}
K.~K. Chang, P.~J. Nair, D.~Lee, S.~Ghose, M.~K. Qureshi, and O.~Mutlu,
  ``{Low-Cost Inter-Linked Subarrays (LISA): Enabling Fast Inter-Subarray Data
  Movement in DRAM},'' in \emph{HPCA}, 2016.

\bibitem{chang.sigmetrics17}
K.~K. Chang, A.~G. Ya{\u g}l{\i}k{\c c}{\i}, S.~Ghose, A.~Agrawal,
  N.~Chatterjee, A.~Kashyap, D.~Lee, M.~O'Connor, H.~Hassan, and O.~Mutlu,
  ``{Understanding Reduced-Voltage Operation in Modern DRAM Devices:
  Experimental Characterization, Analysis, and Mechanisms},'' in
  \emph{SIGMETRICS}, 2017.

\bibitem{cojocar2020susceptible}
L.~Cojocar, J.~Kim, M.~Patel, L.~Tsai, S.~Saroiu, A.~Wolman, and O.~Mutlu,
  ``{Are We Susceptible to Rowhammer? An End-to-End Methodology for Cloud
  Providers},'' in \emph{S\&P}, 2020.

\bibitem{dai2018graphh}
G.~Dai, T.~Huang, Y.~Chi, J.~Zhao, G.~Sun, Y.~Liu, Y.~Wang, Y.~Xie, and
  H.~Yang, ``{GraphH: A Processing-in-Memory Architecture for Large-scale Graph
  Processing},'' in \emph{TCAD}, 2018.

\bibitem{deng.dac2018}
Q.~Deng, L.~Jiang, Y.~Zhang, M.~Zhang, and J.~Yang, ``{DrAcc: a DRAM Based
  Accelerator for Accurate CNN Inference},'' in \emph{DAC}, 2018.

\bibitem{dfi}
{{DFI Group}}, \emph{{{DFI 5.0 Specification}}}, July 2018.

\bibitem{drumond2017mondrian}
M.~P. Drumond Lages De~Oliveira, A.~Daglis, N.~Mirzadeh, D.~Ustiugov,
  J.~Picorel~Obando, B.~Falsafi, B.~Grot, and D.~Pnevmatikatos, ``{The Mondrian
  Data Engine},'' in \emph{ISCA}, 2017.

\bibitem{forlin2022sim2pim}
B.~E.~Forlin, P.~C. Santos, A.~E. Becker, M.~A. Alves, and L.~Carro,
  ``{Sim2PIM: A Complete Simulation Framework for Processing-in-Memory},'' in
  \emph{JSA}, 2022.

\bibitem{eckert2018neural}
C.~Eckert, X.~Wang, J.~Wang, A.~Subramaniyan, R.~Iyer, D.~Sylvester,
  D.~Blaaauw, and R.~Das, ``{Neural Cache: Bit-Serial In-Cache Acceleration of
  Deep Neural Networks},'' in \emph{ISCA}, 2018.

\bibitem{farmahini2015nda}
A.~Farmahini-Farahani, J.~H. Ahn, K.~Morrow, and N.~S. Kim, ``{NDA: Near-DRAM
  Acceleration Architecture Leveraging Commodity DRAM Devices and Standard
  Memory Modules},'' in \emph{HPCA}, 2015.

\bibitem{farmani2021RHAT}
M.~Farmani, M.~Tehranipoor, and F.~Rahman, ``{RHAT: Efficient RowHammer-Aware
  Test for Modern DRAM Modules},'' in \emph{ETS}, 2021.

\bibitem{fernandez2020natsa}
I.~Fernandez, R.~Quislant, C.~Giannoula, M.~Alser, J.~Gomez-Luna, E.~Gutierrez,
  O.~Plata, and O.~Mutlu, ``{NATSA: A Near-Data Processing Accelerator for Time
  Series Analysis},'' in \emph{ICCD}, 2020.

\bibitem{ferreira2021pluto}
J.~D. Ferreira, G.~Falcao, J.~G{\'o}mez-Luna, M.~Alser, L.~Orosa,
  M.~Sadrosadati, J.~S. Kim, G.~F. Oliveira, T.~Shahroodi, A.~Nori
  \emph{et~al.}, ``{pLUTo: In-DRAM Lookup Tables to Enable Massively Parallel
  General-Purpose Computation},'' arXiv:2104.07699, 2021.

\bibitem{frigo2020trr}
P.~Frigo, E.~Vannacci, H.~Hassan, V.~van~der Veen, O.~Mutlu, C.~Giuffrida,
  H.~Bos, and K.~Razavi, ``{TRRespass: Exploiting the Many Sides of Target Row
  Refresh},'' in \emph{S\&P}, 2020.

\bibitem{fujiki2019duality}
D.~Fujiki, S.~Mahlke, and R.~Das, ``{Duality Cache for Data Parallel
  Acceleration},'' in \emph{ISCA}, 2019.

\bibitem{gaillardon2016plim}
P.-E. Gaillardon, L.~Amarú, A.~Siemon, E.~Linn, R.~Waser, A.~Chattopadhyay,
  and G.~De~Micheli, ``{The Programmable Logic-in-Memory (PLiM) Computer},'' in
  \emph{DATE}, 2016.

\bibitem{gao2020computedram}
F.~Gao, G.~Tziantzioulis, and D.~Wentzlaff, ``{ComputeDRAM: In-Memory Compute
  Using Off-the-Shelf DRAMs},'' in \emph{MICRO}, 2019.

\bibitem{gao.pact15}
M.~Gao, G.~Ayers, and C.~Kozyrakis, ``{Practical Near-Data Processing for
  In-Memory Analytics Frameworks},'' in \emph{PACT}, 2015.

\bibitem{DBLP:conf/hpca/GaoK16}
M.~Gao and C.~Kozyrakis, ``{HRL: Efficient and Flexible Reconfigurable Logic
  for Near-Data Processing},'' in \emph{HPCA}, 2016.

\bibitem{gao2017tetris}
M.~Gao, J.~Pu, X.~Yang, M.~Horowitz, and C.~Kozyrakis, ``{Tetris: Scalable and
  Efficient Neural Network Acceleration with 3D Memory},'' in \emph{ASPLOS},
  2017.

\bibitem{ghose2019processing}
S.~Ghose, A.~Boroumand, J.~S. Kim, J.~G{\'o}mez-Luna, and O.~Mutlu,
  ``{Processing-in-Memory: A Workload-driven Perspective},'' in \emph{IBM JRD},
  2019.

\bibitem{ghose2019demystifying}
S.~Ghose, T.~Li, N.~Hajinazar, D.~S. Cali, and O.~Mutlu, ``{Demystifying
  Complex Workload-DRAM Interactions: An Experimental Study},'' in
  \emph{SIGMETRICS}, 2019.

\bibitem{ghose2018vampire}
S.~Ghose, A.~G. Yaglik\c{c}i, R.~Gupta, D.~Lee, K.~Kudrolli, W.~X. Liu,
  H.~Hassan, K.~K. Chang, N.~Chatterjee, A.~Agrawal, M.~O'Connor, and O.~Mutlu,
  ``{What Your DRAM Power Models Are Not Telling You: Lessons from a Detailed
  Experimental Study},'' in \emph{SIGMETRICS}, 2018.

\bibitem{syncron}
C.~Giannoula, N.~Vijaykumar, N.~Papadopoulou, V.~Karakostas, I.~Fernandez,
  J.~Gómez-Luna, L.~Orosa, N.~Koziris, G.~Goumas, and O.~Mutlu, ``{SynCron:
  Efficient Synchronization Support for Near-Data-Processing Architectures},''
  in \emph{HPCA}, 2021.

\bibitem{softmc.github}
\BIBentryALTinterwordspacing
S.~R. Group, ``{SoftMC v1.0 -- GitHub Repository},'' 2021. [Online]. Available:
  \url{https://github.com/CMU-SAFARI/SoftMC}
\BIBentrySTDinterwordspacing

\bibitem{gu.isca16}
B.~Gu, A.~S. Yoon, D.-H. Bae, I.~Jo, J.~Lee, J.~Yoon, J.-U. Kang, M.~Kwon,
  C.~Yoon, S.~Cho, J.~Jeong, and D.~Chang, ``{Biscuit: {A} Framework for
  Near-Data Processing of Big Data Workloads},'' in \emph{ISCA}, 2016.

\bibitem{hajinazarsimdram}
N.~Hajinazar, G.~F. Oliveira, S.~Gregorio, J.~D. Ferreira, N.~M. Ghiasi,
  M.~Patel, M.~Alser, S.~Ghose, J.~G{\'o}mez-Luna, and O.~Mutlu, ``{SIMDRAM: A
  Framework for Bit-Serial SIMD Processing Using DRAM},'' in \emph{ASPLOS},
  2021.

\bibitem{hamdioui2017myth}
S.~Hamdioui, S.~Kvatinsky, and e.~a. G.~Cauwenberghs, ``{Memristor for
  Computing: Myth or Reality?}'' in \emph{DATE}, 2017.

\bibitem{hamdioui2015memristor}
S.~Hamdioui, L.~Xie, H.~A. Du~Nguyen, M.~Taouil, K.~Bertels, H.~Corporaal,
  H.~Jiao, F.~Catthoor, D.~Wouters, L.~Eike, and J.~van Lunteren, ``{Memristor
  Based Computation-in-Memory Architecture for Data-intensive Applications},''
  in \emph{DATE}, 2015.

\bibitem{cont-runahead}
M.~Hashemi, O.~Mutlu, and Y.~N. Patt, ``{Continuous Runahead: Transparent
  Hardware Acceleration for Memory Intensive Workloads},'' in \emph{MICRO},
  2016.

\bibitem{hashemi.isca16}
M.~Hashemi, Khubaib, E.~Ebrahimi, O.~Mutlu, and Y.~N. Patt, ``{Accelerating
  Dependent Cache Misses with an Enhanced Memory Controller},'' in \emph{ISCA},
  2016.

\bibitem{hassan2021uncovering}
H.~Hassan, Y.~C. Tugrul, J.~S. Kim, V.~van~der Veen, K.~Razavi, and O.~Mutlu,
  ``{Uncovering In-DRAM RowHammer Protection Mechanisms: A New Methodology,
  Custom RowHammer Patterns, and Implications},'' arXiv:2110.10603, 2021.

\bibitem{hassan2017softmc}
H.~Hassan, N.~Vijaykumar, S.~Khan, S.~Ghose, K.~Chang, G.~Pekhimenko, D.~Lee,
  O.~Ergin, and O.~Mutlu, ``{SoftMC: A Flexible and Practical Open-Source
  Infrastructure for Enabling Experimental DRAM Studies},'' in \emph{HPCA},
  2017.

\bibitem{helm2020Reliable}
C.~Helm, S.~Akiyama, and K.~Taura, ``Reliable {{Reverse Engineering}} of
  {{Intel DRAM Addressing Using Performance Counters}},'' in \emph{MASCOTS},
  2020.

\bibitem{hillenbrand2017Physical}
M.~Hillenbrand, ``{Physical Address Decoding in Intel Xeon v3/v4 CPUs: A
  Supplemental Datasheet},'' 2017.

\bibitem{horiguchi1997redundancy}
M.~Horiguchi, ``{Redundancy Techniques for High-Density DRAMs},'' in
  \emph{ISIS}, 1997.

\bibitem{scarab}
\BIBentryALTinterwordspacing
{HPS Research Group}, ``{Scarab -- Github Repository},'' 2022. [Online].
  Available: \url{https://github.com/hpsresearchgroup/scarab}
\BIBentrySTDinterwordspacing

\bibitem{impica}
K.~Hsieh, S.~Khan, N.~Vijaykumar, K.~K. Chang, A.~Boroumand, S.~Ghose, and
  O.~Mutlu, ``{Accelerating Pointer Chasing in 3D-Stacked Memory: Challenges,
  Mechanisms, Evaluation},'' in \emph{ICCD}, 2016.

\bibitem{hsieh.isca16}
K.~Hsieh, E.~Ebrahimi, G.~Kim, N.~Chatterjee, M.~O'Conner, N.~Vijaykumar,
  O.~Mutlu, and S.~Keckler, ``{Transparent Offloading and Mapping (TOM):
  Enabling Programmer-Transparent Near-Data Processing in GPU Systems},'' in
  \emph{ISCA}, 2016.

\bibitem{huang2020heterogeneous}
Y.~Huang, L.~Zheng, P.~Yao, J.~Zhao, X.~Liao, H.~Jin, and J.~Xue, ``{A
  Heterogeneous PIM Hardware-Software Co-Design for Energy-Efficient Graph
  Processing},'' in \emph{IPDPS}, 2020.

\bibitem{x86-manual}
\BIBentryALTinterwordspacing
Intel, ``{Intel 64 and IA-32 Architectures Software Developer Manuals},'' 2011.
  [Online]. Available:
  \url{http://www.intel.com/content/www/us/en/processors/architectures-software-developer-manuals.html}
\BIBentrySTDinterwordspacing

\bibitem{intel-loihi}
Intel, ``{Taking Neuromorphic Computing to the Next Level with Loihi 2},''
  Technology Brief, 2022.

\bibitem{itoh2013vlsi}
K.~Itoh, \emph{{VLSI Memory Chip Design}}.\hskip 1em plus 0.5em minus
  0.4em\relax Springer, 2001.

\bibitem{jedecDDR4}
JEDEC, ``{DDR4},'' \emph{JEDEC Standard JESD79--4}, 2012.

\bibitem{kang2009one}
H.~B. Kang and S.~K. Hong, ``{One-Transistor Type DRAM},'' US Patent 7701751,
  2009.

\bibitem{kang.icassp14}
M.~Kang, M.-S. Keel, N.~R. Shanbhag, S.~Eilert, and K.~Curewitz, ``{An
  Energy-Efficient VLSI Architecture for Pattern Recognition via Deep Embedding
  of Computation in SRAM},'' in \emph{ICASSP}, 2014.

\bibitem{ke2021near}
L.~Ke, X.~Zhang, J.~So, J.-G. Lee, S.-H. Kang, S.~Lee, S.~Han, Y.~Cho, J.~H.
  Kim, Y.~Kwon \emph{et~al.}, ``{Near-Memory Processing in Action: Accelerating
  Personalized Recommendation with AxDIMM},'' in \emph{IEEE Micro}, 2021.

\bibitem{keeth2001dram}
B.~Keeth and R.~Baker, \emph{{DRAM Circuit Design: A Tutorial}}.\hskip 1em plus
  0.5em minus 0.4em\relax Wiley, 2001.

\bibitem{khan.dsn16}
S.~Khan, D.~Lee, and O.~Mutlu, ``{PARBOR: An Efficient System-Level Technique
  to Detect Data Dependent Failures in DRAM},'' in \emph{DSN}, 2016.

\bibitem{khan.micro17}
S.~Khan, C.~Wilkerson, Z.~Wang, A.~Alameldeen, D.~Lee, and O.~Mutlu,
  ``{Detecting and Mitigating Data-Dependent DRAM Failures by Exploiting
  Current Memory Content},'' in \emph{MICRO}, 2017.

\bibitem{kim2021SilentPIM}
C.~H. Kim, W.~J. Lee, Y.~Paik, K.~Kwon, S.~Y. Kim, I.~Park, and S.~W. Kim,
  ``Silent-{{PIM}}: {{Realizing}} the {{Processing}}-in-{{Memory Computing}}
  with {{Standard Memory Requests}},'' \emph{TPDS}, 2021.

\bibitem{kim.isca16}
D.~Kim, J.~Kung, S.~Chai, S.~Yalamanchili, and S.~Mukhopadhyay, ``{Neurocube:
  {A} Programmable Digital Neuromorphic Architecture with High-Density {3D}
  Memory},'' in \emph{ISCA}, 2016.

\bibitem{kim.sc17}
G.~Kim, N.~Chatterjee, M.~O'Connor, and K.~Hsieh, ``{Toward Standardized
  Near-Data Processing with Unrestricted Data Placement for GPUs},'' in
  \emph{SC}, 2017.

\bibitem{kim2018solar}
J.~{Kim}, M.~{Patel}, H.~{Hassan}, and O.~{Mutlu}, ``{Solar-DRAM: Reducing DRAM
  Access Latency by Exploiting the Variation in Local Bitlines},'' in
  \emph{ICCD}, 2018.

\bibitem{kim.hpca18}
J.~Kim, M.~Patel, H.~Hassan, and O.~Mutlu, ``{The {DRAM} Latency {PUF}: Quickly
  Evaluating Physical Unclonable Functions by Exploiting the
  Latency--Reliability Tradeoff in Modern {DRAM} Devices},'' in \emph{HPCA},
  2018.

\bibitem{kim.hpca19}
J.~Kim, M.~Patel, H.~Hassan, L.~Orosa, and O.~Mutlu, ``{D-RaNGe: Using
  Commodity {DRAM} Devices to Generate True Random Numbers with Low Latency and
  High Throughput},'' in \emph{HPCA}, 2019.

\bibitem{kim.bmc18}
J.~S. Kim, D.~Senol, H.~Xin, D.~Lee, S.~Ghose, M.~Alser, H.~Hassan, O.~Ergin,
  C.~Alkan, and O.~Mutlu, ``{GRIM-Filter: Fast Seed Location Filtering in DNA
  Read Mapping Using Processing-in-Memory Technologies},'' in \emph{BMC
  Genomics}, 2018.

\bibitem{kim2020revisiting}
J.~S. Kim, M.~Patel, A.~G. Ya\u{g}l\i{}k\c{c}\i{}, H.~Hassan, R.~Azizi,
  L.~Orosa, and O.~Mutlu, ``{Revisiting RowHammer: An Experimental Analysis of
  Modern DRAM Devices and Mitigation Techniques},'' in \emph{ISCA}, 2020.

\bibitem{kim2021aquabolt}
J.~H. Kim, S.-h. Kang, S.~Lee, H.~Kim, W.~Song, Y.~Ro, S.~Lee, D.~Wang,
  H.~Shin, B.~Phuah \emph{et~al.}, ``{Aquabolt-XL: Samsung HBM2-PIM with
  in-memory processing for ML accelerators and beyond},'' in \emph{Hot Chips},
  2021.

\bibitem{yoongu-thesis}
Y.~Kim, ``{Architectural Techniques to Enhance DRAM Scaling},'' Ph.D.
  dissertation, Carnegie Mellon University, 2015.

\bibitem{kim-isca2014}
Y.~Kim, R.~Daly, J.~Kim, C.~Fallin, J.~H. Lee, D.~Lee, C.~Wilkerson, K.~Lai,
  and O.~Mutlu, ``{Flipping Bits in Memory Without Accessing Them: An
  Experimental Study of DRAM Disturbance Errors},'' in \emph{ISCA}, 2014.

\bibitem{salp}
Y.~Kim, V.~Seshadri, D.~Lee, J.~Liu, and O.~Mutlu, ``{A Case for Exploiting
  Subarray-Level Parallelism (SALP) in DRAM},'' in \emph{ISCA}, 2012.

\bibitem{ramulator}
Y.~Kim, W.~Yang, and O.~Mutlu, ``{Ramulator: A Fast and Extensible DRAM
  Simulator},'' in \emph{CAL}, 2015.

\bibitem{kvatinsky.tcasii14}
S.~Kvatinsky, D.~Belousov, S.~Liman, G.~Satat, N.~Wald, E.~G. Friedman,
  A.~Kolodny, and U.~C. Weiser, ``{MAGIC---Memristor-Aided Logic},'' in
  \emph{IEEE TCAS II: Express Briefs}, 2014.

\bibitem{kvatinsky.iccd11}
S.~Kvatinsky, A.~Kolodny, U.~C. Weiser, and E.~G. Friedman, ``{Memristor-Based
  IMPLY Logic Design Procedure},'' in \emph{ICCD}, 2011.

\bibitem{kvatinsky.tvlsi14}
S.~Kvatinsky, G.~Satat, N.~Wald, E.~G. Friedman, A.~Kolodny, and U.~C. Weiser,
  ``{Memristor-Based Material Implication (IMPLY) Logic: Design Principles and
  Methodologies},'' in \emph{TVLSI}, 2014.

\bibitem{kwon2021fimdram}
Y.-C. Kwon, S.~H. Lee, J.~Lee, S.-H. Kwon, J.~M. Ryu, J.-P. Son, O.~Seongil,
  H.-S. Yu, H.~Lee, S.~Y. Kim, Y.~Cho, J.~G. Kim, J.~Choi, H.-S. Shin, J.~Kim,
  B.~Phuah, H.~Kim, M.~J. Song, A.~Choi, D.~Kim, S.~Kim, E.-B. Kim, D.~Wang,
  S.~Kang, Y.~Ro, S.~Seo, J.~Song, J.~Youn, K.~Sohn, and N.~S. Kim, ``25.4
  {{A}} 20nm {{6GB Function}}-{{In}}-{{Memory DRAM}}, {{Based}} on {{HBM2}}
  with a 1.{{2TFLOPS Programmable Computing Unit Using Bank}}-{{Level
  Parallelism}}, for {{Machine Learning Applications}},'' in \emph{ISSCC},
  2021.

\bibitem{lee.thesis16}
D.~Lee, ``{Reducing DRAM Latency at Low Cost by Exploiting Heterogeneity},''
  Ph.D. dissertation, Carnegie Mellon University, 2016.

\bibitem{lee.sigmetrics17}
D.~Lee, S.~Khan, L.~Subramanian, S.~Ghose, R.~Ausavarungnirun, G.~Pekhimenko,
  V.~Seshadri, and O.~Mutlu, ``{Design-Induced Latency Variation in Modern DRAM
  Chips: Characterization, Analysis, and Latency Reduction Mechanisms},'' in
  \emph{{SIGMETRICS}}, 2017.

\bibitem{lee2022improving}
D.~Lee, J.~So, M.~AHN, J.-G. Lee, J.~Kim, J.~Cho, R.~Oliver, V.~C. Thummala,
  R.~s. JV, S.~S. Upadhya \emph{et~al.}, ``{Improving In-Memory Database
  Operations with Acceleration DIMM (AxDIMM)},'' in \emph{DaMoN}, 2022.

\bibitem{lee.hpca15}
D.~Lee, Y.~Kim, G.~Pekhimenko, S.~Khan, V.~Seshadri, K.~Chang, and O.~Mutlu,
  ``{Adaptive-Latency DRAM: Optimizing DRAM Timing for the Common-Case},'' in
  \emph{HPCA}, 2015.

\bibitem{lee.hpca13}
D.~Lee, Y.~Kim, V.~Seshadri, J.~Liu, L.~Subramanian, and O.~Mutlu,
  ``{Tiered-Latency DRAM: A Low Latency and Low Cost DRAM Architecture},'' in
  \emph{HPCA}, 2013.

\bibitem{donghyuk-ddma}
D.~Lee, L.~Subramanian, R.~Ausavarungnirun, J.~Choi, and O.~Mutlu, ``{Decoupled
  Direct Memory Access: Isolating CPU and IO Traffic by Leveraging a
  Dual-Data-Port DRAM},'' in \emph{PACT}, 2015.

\bibitem{lee2022isscc}
S.~Lee, K.~Kim, S.~Oh, J.~Park, G.~Hong, D.~Ka, K.~Hwang, J.~Park, K.~Kang,
  J.~Kim, J.~Jeon, N.~Kim, Y.~Kwon, K.~Vladimir, W.~Shin, J.~Won, M.~Lee,
  H.~Joo, H.~Choi, J.~Lee, D.~Ko, Y.~Jun, K.~Cho, I.~Kim, C.~Song, C.~Jeong,
  D.~Kwon, J.~Jang, I.~Park, J.~Chun, and J.~Cho, ``{A 1ynm 1.25V 8Gb,
  16Gb/s/pin GDDR6-based Accelerator-in-Memory supporting 1TFLOPS MAC Operation
  and Various Activation Functions for Deep-Learning Applications},'' in
  \emph{ISSCC}, 2022.

\bibitem{levy.microelec14}
Y.~Levy, J.~Bruck, Y.~Cassuto, E.~G. Friedman, A.~Kolodny, E.~Yaakobi, and
  S.~Kvatinsky, ``{Logic Operations in Memory Using a Memristive Akers
  Array},'' in \emph{Microelectronics Journal}, 2014.

\bibitem{li.micro17}
S.~Li, D.~Niu, K.~T. Malladi, H.~Zheng, B.~Brennan, and Y.~Xie, ``{DRISA: A
  DRAM-Based Reconfigurable In-Situ Accelerator},'' in \emph{MICRO}, 2017.

\bibitem{li.dac16}
S.~Li, C.~Xu, Q.~Zou, J.~Zhao, Y.~Lu, and Y.~Xie, ``{Pinatubo: A
  Processing-in-Memory Architecture for Bulk Bitwise Operations in Emerging
  Non-Volatile Memories},'' in \emph{DAC}, 2016.

\bibitem{calloc}
{Linux man-pages Project}, ``{calloc(3p) — Linux manual page},''
  \url{https://man7.org/linux/man-pages/man3/calloc.3p.html}, 2022.

\bibitem{malloc}
{Linux man-pages Project}, ``{malloc(3) — Linux manual page},''
  \url{https://man7.org/linux/man-pages/man3/malloc.3.html}, 2022.

\bibitem{memcpy}
{Linux man-pages Project}, ``{memcpy(3) — Linux manual page},''
  \url{https://man7.org/linux/man-pages/man3/memcpy.3.html}, 2022.

\bibitem{posixmemalign}
{Linux man-pages Project}, ``{posix\_memalign(3) — Linux manual page},''
  \url{https://man7.org/linux/man-pages/man3/posix_memalign.3.html}, 2022.

\bibitem{perfLinux}
{Linux Wiki}, ``{perf: Linux profiling with performance counters},''
  \url{https://perf.wiki.kernel.org/index.php/Main_Page}, 2021.

\bibitem{liu.isca13}
J.~Liu, B.~Jaiyen, Y.~Kim, C.~Wilkerson, and O.~Mutlu, ``{An Experimental Study
  of Data Retention Behavior in Modern {DRAM} Devices: Implications for
  Retention Time Profiling Mechanisms},'' in \emph{ISCA}, 2013.

\bibitem{liu-spaa17}
Z.~Liu, I.~Calciu, M.~Herlihy, and O.~Mutlu, ``{Concurrent Data Structures for
  Near-Memory Computing},'' in \emph{SPAA}, 2017.

\bibitem{lu2015improving}
S.-L. Lu, Y.-C. Lin, and C.-L. Yang, ``{Improving DRAM Latency with Dynamic
  Asymmetric Subarray},'' in \emph{MICRO}, 2015.

\bibitem{luo2020clr}
H.~Luo, T.~Shahroodi, H.~Hassan, M.~Patel, A.~G. Yaglikci, L.~Orosa, J.~Park,
  and O.~Mutlu, ``{CLR-DRAM: A Low-Cost DRAM Architecture Enabling Dynamic
  Capacity-Latency Trade-Off},'' in \emph{ISCA}, 2020.

\bibitem{mandelman.ibmjrd02}
J.~A. Mandelman, R.~H. Dennard, G.~B. Bronner, J.~K. DeBrosse, R.~Divakaruni,
  Y.~Li, and C.~J. Radens, ``{Challenges and Future Directions for the Scaling
  of Dynamic Random-Access Memory ({DRAM})},'' in \emph{IBM JRD}, 2002.

\bibitem{maxwellFT200}
{Maxwell}, ``{FT20X},''
  \url{https://www.maxwell-fa.com/upload/files/base/8/m/311.pdf}, 2022.

\bibitem{micron2016ddr4}
Micron, ``{DDR4 SDRAM Datasheet},'' 2016.

\bibitem{micron2018ddr3}
Micron, ``{DDR3 SDRAM: MT41J128M8},'' Data Sheet, 2018.

\bibitem{morad.taco15}
A.~Morad, L.~Yavits, and R.~Ginosar, ``{GP-SIMD Processing-in-Memory},'' in
  \emph{ACM TACO}, 2015.

\bibitem{mosanu2022pimulator}
S.~Mosanu, M.~N. Sakib, T.~II, E.~Cukurtas, A.~Ahmed, P.~Ivanov, S.~Khan,
  K.~Skadron, and M.~Stan, ``{PiMulator: a Fast and Flexible
  Processing-in-Memory Emulation Platform},'' in \emph{DATE}, 2022.

\bibitem{mutlu2020modern}
O.~Mutlu, S.~Ghose, J.~G{\'o}mez-Luna, and R.~Ausavarungnirun, ``{A Modern
  Primer on Processing in Memory},'' in \emph{Emerging Computing: From Devices
  to Systems - Looking Beyond Moore and Von Neumann}, 2021.

\bibitem{nai2017graphpim}
L.~Nai, R.~Hadidi, J.~Sim, H.~Kim, P.~Kumar, and H.~Kim, ``{GraphPIM: Enabling
  Instruction-Level PIM Offloading in Graph Computing Frameworks},'' in
  \emph{HPCA}, 2017.

\bibitem{niu2022isscc}
D.~Niu, S.~Li, Y.~Wang, W.~Han, Z.~Zhang, Y.~Guan, T.~Guan, F.~Sun, F.~Xue,
  L.~Duan \emph{et~al.}, ``{184QPS/W 64Mb/mm 2 3D Logic-to-DRAM Hybrid Bonding
  with Process-Near-Memory Engine for Recommendation System},'' in
  \emph{ISSCC}, 2022.

\bibitem{olgun2021quactrng}
A.~Olgun, M.~Patel, A.~G. Yaglikci, H.~Luo, J.~S. Kim, N.~Bostanci,
  N.~Vijaykumar, O.~Ergin, and O.~Mutlu, ``{QUAC-TRNG: High-Throughput True
  Random Number Generation Using Quadruple Row Activation in Commodity DRAM
  Chips},'' {arXiv:2105.08955}, 2021.

\bibitem{olgun2021quactrngieee}
A.~Olgun, M.~Patel, A.~G. Ya\u{g}l\i{}k\c{c}\i{}, H.~Luo, J.~S. Kim,
  F.~Nisa~Bostancı, N.~Vijaykumar, O.~Ergin, and O.~Mutlu, ``{QUAC-TRNG:
  High-Throughput True Random Number Generation Using Quadruple Row Activation
  in Commodity DRAM Chips},'' in \emph{ISCA}, 2021.

\bibitem{oliveira2021damov}
G.~F. Oliveira, J.~Gómez-Luna, L.~Orosa, S.~Ghose, N.~Vijaykumar,
  I.~Fernandez, M.~Sadrosadati, and O.~Mutlu, ``{DAMOV: A New Methodology and
  Benchmark Suite for Evaluating Data Movement Bottlenecks},'' in \emph{IEEE
  Access}, 2021.

\bibitem{orosa2021codic}
L.~Orosa, Y.~Wang, M.~Sadrosadati, J.~S. Kim, M.~Patel, I.~Puddu, H.~Luo,
  K.~Razavi, J.~Gómez-Luna, H.~Hassan, N.~Mansouri-Ghiasi, S.~Ghose, and
  O.~Mutlu, ``{CODIC: A Low-Cost Substrate for Enabling Custom In-DRAM
  Functionalities and Optimizations},'' in \emph{ISCA}, 2021.

\bibitem{orosaYaglikci2021deeper}
L.~Orosa, A.~G. Yaglikci, H.~Luo, A.~Olgun, J.~Park, H.~Hassan, M.~Patel, J.~S.
  Kim, and O.~Mutlu, ``{A Deeper Look into RowHammer’s Sensitivities:
  Experimental Analysis of Real DRAM Chipsand Implications on Future Attacks
  and Defenses},'' in \emph{MICRO}, 2021.

\bibitem{patel2017reaper}
M.~Patel, J.~S. Kim, and O.~Mutlu, ``{The Reach Profiler (REAPER): Enabling the
  Mitigation of DRAM Retention Failures via Profiling at Aggressive
  Conditions},'' in \emph{ISCA}, 2017.

\bibitem{patel2020bit}
M.~Patel, J.~S. Kim, T.~Shahroodi, H.~Hassan, and O.~Mutlu, ``{Bit-Exact ECC
  Recovery (BEER): Determining DRAM On-Die ECC Functions by Exploiting DRAM
  Data Retention Characteristics},'' in \emph{MICRO}, 2020.

\bibitem{patel2022case}
M.~Patel, T.~Shahroodi, A.~Manglik, A.~G. Yaglikci, A.~Olgun, H.~Luo, and
  O.~Mutlu, ``{A Case for Transparent Reliability in DRAM Systems},''
  arXiv:2204.10378, 2022.

\bibitem{pattnaik.pact16}
A.~Pattnaik, X.~Tang, A.~Jog, O.~Kayiran, A.~K. Mishra, M.~T. Kandemir,
  O.~Mutlu, and C.~R. Das, ``{Scheduling Techniques for GPU Architectures with
  Processing-in-Memory Capabilities},'' in \emph{PACT}, 2016.

\bibitem{gem5-gpu}
J.~Power, J.~Hestness, M.~S. Orr, M.~D. Hill, and D.~A. Wood, ``{gem5-gpu: A
  Heterogeneous CPU-GPU Simulator},'' in \emph{CAL}, Jan 2015.

\bibitem{pugsley2014ndc}
S.~H. Pugsley, J.~Jestes, H.~Zhang, R.~Balasubramonian, V.~Srinivasan,
  A.~Buyuktosunoglu, A.~Davis, and F.~Li, ``{{NDC: Analyzing the Impact of
  3D-Stacked Memory+Logic Devices on MapReduce Workloads}},'' in \emph{ISPASS},
  2014.

\bibitem{rezaei2020nom}
S.~H.~S. {Rezaei}, M.~{Modarressi}, R.~{Ausavarungnirun}, M.~{Sadrosadati},
  O.~{Mutlu}, and M.~{Daneshtalab}, ``{NoM: Network-on-Memory for Inter-Bank
  Data Transfer in Highly-Banked Memories},'' in \emph{CAL}, 2020.

\bibitem{riscv-gnu-toolchain}
\BIBentryALTinterwordspacing
RISC-V, ``{RISC-V GNU Compiler Toolchain},'' 2021. [Online]. Available:
  \url{https://github.com/riscv/riscv-gnu-toolchain}
\BIBentrySTDinterwordspacing

\bibitem{riscv-pk}
\BIBentryALTinterwordspacing
RISC-V, ``{RISC-V} proxy kernel,'' 2022. [Online]. Available:
  \url{https://github.com/riscv/riscv-pk}
\BIBentrySTDinterwordspacing

\bibitem{ronen2022bitlet}
R.~Ronen, A.~Eliahu, O.~Leitersdorf, N.~Peled, K.~Korgaonkar, A.~Chattopadhyay,
  B.~Perach, and S.~Kvatinsky, ``{The Bitlet Model: A Parameterized Analytical
  Model to Compare PIM and CPU Systems},'' in \emph{J. Emerg. Technol. Comput.
  Syst.}, 2022.

\bibitem{ramulator.github}
{SAFARI Research Group}, ``{Ramulator: A DRAM Simulator -- GitHub
  Repository},'' \url{https://github.com/CMU-SAFARI/ramulator/}, 2015.

\bibitem{geraldodamov}
{SAFARI Research Group}, ``{DAMOV -- GitHub Repository},''
  \url{https://github.com/CMU-SAFARI/DAMOV}, 2021.

\bibitem{ramulator-pim}
\BIBentryALTinterwordspacing
{SAFARI Research Group}, ``{Ramulator-PIM: A Processing-in-Memory Simulation
  Framework -- GitHub Repository},'' 2021. [Online]. Available:
  \url{https://github.com/CMU-SAFARI/ramulator-pim}
\BIBentrySTDinterwordspacing

\bibitem{zsim}
D.~Sanchez and C.~Kozyrakis, ``{ZSim: Fast and Accurate Microarchitectural
  Simulation of Thousand-Core Systems},'' in \emph{ISCA}, 2013.

\bibitem{saroiu2022price}
S.~Saroiu, A.~Wolman, and L.~Cojocar, ``{The Price of Secrecy: How Hiding
  Internal DRAM Topologies Hurts Rowhammer Defenses},'' in \emph{IRPS}, 2022.

\bibitem{seshadri.thesis16}
V.~Seshadri, ``{Simple DRAM and Virtual Memory Abstractions to Enable Highly
  Efficient Memory Systems},'' Ph.D. dissertation, Carnegie Mellon University,
  2016.

\bibitem{seshadri.arxiv16}
V.~Seshadri, D.~Lee, T.~Mullins, H.~Hassan, A.~Boroumand, J.~Kim, M.~A. Kozuch,
  O.~Mutlu, P.~B. Gibbons, and T.~C. Mowry, ``{Buddy-RAM: Improving the
  Performance and Efficiency of Bulk Bitwise Operations Using DRAM},''
  arXiv:1611.09988, 2016.

\bibitem{seshadri.micro17}
V.~Seshadri, D.~Lee, T.~Mullins, H.~Hassan, A.~Boroumand, J.~Kim, M.~A. Kozuch,
  O.~Mutlu, P.~B. Gibbons, and T.~C. Mowry, ``{Ambit: In-Memory Accelerator for
  Bulk Bitwise Operations Using Commodity DRAM Technology},'' in \emph{MICRO},
  2017.

\bibitem{seshadri2014dirty}
V.~Seshadri, A.~Bhowmick, O.~Mutlu, P.~B. Gibbons, M.~A. Kozuch, and T.~C.
  Mowry, ``{The Dirty-Block Index},'' in \emph{ISCA}, 2014.

\bibitem{Seshadri:2015:ANDOR}
V.~Seshadri, K.~Hsieh, A.~Boroumand, D.~Lee, M.~A. Kozuch, O.~Mutlu, P.~B.
  Gibbons, and T.~C. Mowry, ``{Fast Bulk Bitwise AND and OR in DRAM},'' in
  \emph{CAL}, 2015.

\bibitem{seshadri2013rowclone}
V.~Seshadri, Y.~Kim, C.~Fallin, D.~Lee, R.~Ausavarungnirun, G.~Pekhimenko,
  Y.~Luo, O.~Mutlu, M.~A. Kozuch, P.~B. Gibbons, and T.~C. Mowry, ``{RowClone:
  Fast and Energy-Efficient In-DRAM Bulk Data Copy and Initialization},'' in
  \emph{MICRO}, 2013.

\bibitem{seshadri.bookchapter17.arxiv}
V.~Seshadri and O.~Mutlu, ``{The Processing Using Memory Paradigm: In-DRAM Bulk
  Copy, Initialization, Bitwise AND and OR},'' arXiv:1610.09603, 2016.

\bibitem{seshadri.bookchapter17}
V.~Seshadri and O.~Mutlu, ``{Simple Operations in Memory to Reduce Data
  Movement},'' in \emph{Advances in Computers, Volume 106}, 2017.

\bibitem{seshadri2020indram}
V.~Seshadri and O.~Mutlu, ``{In-DRAM Bulk Bitwise Execution Engine},''
  arXiv:1905.09822, 2020.

\bibitem{shafiee2016isaac}
A.~Shafiee, A.~Nag, N.~Muralimanohar, R.~Balasubramonian, J.~P. Strachan,
  M.~Hu, R.~S. Williams, and V.~Srikumar, ``{ISAAC: A Convolutional Neural
  Network Accelerator with In-situ Analog Arithmetic in Crossbars},'' in
  \emph{ISCA}, 2016.

\bibitem{singh2019napel}
G.~Singh, J.~Gomez-Luna, G.~Mariani, G.~F. Oliveira, S.~Corda, S.~Stujik,
  O.~Mutlu, and H.~Corporaal, ``{NAPEL: Near-memory Computing Application
  Performance Prediction via Ensemble Learning},'' in \emph{DAC}, 2019.

\bibitem{spec2006}
\BIBentryALTinterwordspacing
{Standard Performance Evaluation Corp.}, ``{SPEC CPU 2006},'' 2006. [Online].
  Available: \url{http://www.spec.org/cpu2006}
\BIBentrySTDinterwordspacing

\bibitem{talukder2020towards}
B.~S.~B. Talukder, V.~Menon, B.~Ray, T.~Neal, and M.~Rahman, ``{Towards the
  Avoidance of Counterfeit Memory: Identifying the DRAM Origin},'' in
  \emph{HOST}, 2020.

\bibitem{testa2016inversion}
E.~Testa, M.~Soeken, O.~Zografos, L.~Amaru, P.~Raghavan, R.~Lauwereins, P.-E.
  Gaillardon, and G.~De~Micheli, ``{Inversion Optimization in Majority-Inverter
  Graphs},'' in \emph{NANOARCH}, 2016.

\bibitem{pl068p}
TTi, ``{PL \& PL-P Series DC Power Supplies Data Sheet - Issue 5},''
  \url{https://resources.aimtti.com/datasheets/AIM-PL+PL-P_series_DC_power_supplies_data_sheet-Iss5.pdf},
  2022.

\bibitem{upmem2018}
UPMEM, ``{Introduction to {UPMEM PIM}. Processing-in-memory {(PIM)} on {DRAM}
  Accelerator},'' 2018.

\bibitem{vandegoor2002address}
A.~van~de Goor and I.~Schanstra, ``Address and data scrambling: Causes and
  impact on memory tests,'' in \emph{IEEE International Workshop on Electronic
  Design, Test and Applications}, 2002.

\bibitem{wang2020figaro}
Y.~Wang, L.~Orosa, X.~Peng, Y.~Guo, S.~Ghose, M.~Patel, J.~S. Kim, J.~G. Luna,
  M.~Sadrosadati, N.~M. Ghiasi, and O.~Mutlu, ``{FIGARO: Improving System
  Performance via Fine-Grained In-DRAM Data Relocation and Caching},'' in
  \emph{MICRO}, 2020.

\bibitem{riscv-spec}
\BIBentryALTinterwordspacing
A.~Waterman and K.~Asanovic, ``{The RISC-V Instruction Set Manual},'' 2021.
  [Online]. Available:
  \url{https://riscv.org/wp-content/uploads/2019/06/riscv-spec.pdf}
\BIBentrySTDinterwordspacing

\bibitem{JAFAR}
S.~L. Xi, O.~Babarinsa, M.~Athanassoulis, and S.~Idreos, ``{Beyond the Wall:
  Near-Data Processing for Databases},'' in \emph{DaMoN}, 2015.

\bibitem{xie2015fast}
L.~Xie, H.~A.~D. Nguyen, M.~Taouil, S.~Hamdioui, and K.~Bertels, ``{Fast
  Boolean Logic Mapped on Memristor Crossbar},'' in \emph{ICCD}, 2015.

\bibitem{virtex7mig}
Xilinx, \emph{{7 Series FPGAs Memory Interface Solutions}}, March 2011.

\bibitem{designconstraints}
Xilinx, \emph{{Vivado Design Suite: Using Constraints}}, November 2021.

\bibitem{zynq}
\BIBentryALTinterwordspacing
Xilinx, ``{Xilinx Ultrascale+ MPSoC},'' 2021. [Online]. Available:
  \url{https://www.xilinx.com/products/silicon-devices/soc/zynq-ultrascale-mpsoc.html}
\BIBentrySTDinterwordspacing

\bibitem{zc706}
\BIBentryALTinterwordspacing
Xilinx, ``{Xilinx Zynq-7000 SoC ZC706 Evaluation Kit},'' 2021. [Online].
  Available:
  \url{https://www.xilinx.com/products/boards-and-kits/ek-z7-zc706-g.html}
\BIBentrySTDinterwordspacing

\bibitem{xin2020elp2im}
X.~Xin, Y.~Zhang, and J.~Yang, ``{ELP2IM: Efficient and Low Power Bitwise
  Operation Processing in DRAM},'' in \emph{HPCA}, 2020.

\bibitem{xu2019pimsim}
S.~Xu, X.~Chen, Y.~Wang, Y.~Han, X.~Qian, and X.~Li, ``{PIMSim: A Flexible and
  Detailed Processing-in-Memory Simulator},'' in \emph{IEEE CAL}, 2019.

\bibitem{yu2021multipim}
C.~Yu, S.~Liu, and S.~Khan, ``{MultiPIM: A Detailed and Configurable
  Multi-Stack Processing-In-Memory Simulator},'' in \emph{IEEE CAL}, 2021.

\bibitem{yu2018memristive}
J.~Yu, H.~A.~D. Nguyen, L.~Xie, M.~Taouil, and S.~Hamdioui, ``{Memristive
  Devices for Computation-in-Memory},'' in \emph{DATE}, 2018.

\bibitem{zha2019liquid}
Y.~Zha, E.~Nowak, and J.~Li, ``{Liquid Silicon: A Nonvolatile Fully
  Programmable Processing-In-Memory Processor with Monolithically Integrated
  ReRAM for Big Data/Machine Learning Applications},'' in \emph{Symposium on
  VLSI Circuits}, 2019.

\bibitem{zhang.hpdc14}
D.~P. Zhang, N.~Jayasena, A.~Lyashevsky, J.~L. Greathouse, L.~Xu, and
  M.~Ignatowski, ``{TOP-PIM: Throughput-Oriented Programmable Processing in
  Memory},'' in \emph{HPDC}, 2014.

\bibitem{zhang2020MEG}
J.~Zhang, Y.~Zha, N.~Beckwith, B.~Liu, and J.~Li, ``{{MEG}}: {{A RISCV}}-based
  {{System Emulation Infrastructure}} for {{Near}}-data {{Processing Using
  FPGAs}} and {{High}}-bandwidth {{Memory}},'' in \emph{TRETS}, 2020.

\bibitem{zhang2022pim}
L.~Zhang and L.~Shen, ``{PIM-HBMSim: A Processing in Memory Simulator Based on
  High Bandwidth Memory},'' in \emph{CICA}, 2022.

\bibitem{zhang2018graphp}
M.~Zhang, Y.~Zhuo, C.~Wang, M.~Gao, Y.~Wu, K.~Chen, C.~Kozyrakis, and X.~Qian,
  ``{GraphP: Reducing Communication for PIM-based Graph Processing with
  Efficient Data Partition},'' in \emph{HPCA}, 2018.

\bibitem{zhu2013accelerating}
Q.~Zhu, T.~Graf, H.~E. Sumbul, L.~Pileggi, and F.~Franchetti, ``{Accelerating
  Sparse Matrix-Matrix Multiplication with 3D-Stacked Logic-in-Memory
  Hardware},'' in \emph{HPEC}, 2013.

\bibitem{zhuo2019graphq}
Y.~Zhuo, C.~Wang, M.~Zhang, R.~Wang, D.~Niu, Y.~Wang, and X.~Qian, ``{GraphQ:
  Scalable PIM-based Graph Processing},'' in \emph{MICRO}, 2019.

\end{thebibliography}
\end{footnotesize}
\end{spacing}
\newpage

\end{document}